# On Non-coherent MIMO Channels in the Wideband Regime: Capacity and Reliability *

Siddharth Ray† Muriel Médard† Lizhong Zheng†

**Abstract**

We consider a multiple-input, multiple-output (MIMO) wideband Rayleigh block fading channel where the channel state is unknown to both the transmitter and the receiver and there is only an average power constraint on the input. We compute the capacity and analyze its dependence on coherence length, number of antennas and receive signal-to-noise ratio (SNR) per degree of freedom. We establish conditions on the coherence length and number of antennas for the non-coherent channel to have a "near coherent" performance in the wideband regime. We also propose a signaling scheme that is near-capacity achieving in this regime.

We compute the error probability for this wideband non-coherent MIMO channel and study its dependence on SNR, number of transmit and receive antennas and coherence length. We show that error probability decays inversely with coherence length and exponentially with the product of the number of transmit and receive antennas. Moreover, channel outage dominates error probability in the wideband regime. We also show that the critical as well as cut-off rates are much smaller than channel capacity in this regime.

# 1 Introduction

Recent years have seen the emergence of high data rate, third generation wideband wireless communication standards like wideband code division multiple access (W-CDMA) and Ultra-wideband (UWB) radio. Motivated by the ever increasing demand for higher wideband wireless data rates, we consider multiple antenna communication over the wideband wireless channel.

At the cost of additional signal processing (which is getting cheaper with rapid advances in VLSI technology), multiple-input, multiple-output (MIMO) systems have been known to improve considerably performance of wireless systems in terms of reliability as well as throughput, without requiring additional resources such as bandwidth and power. However, multiple antenna research has focused primarily in the regime where the received signal-to-noise ratio (SNR) per degree of freedom is high. Such a regime operates in essence as a narrowband regime. We now study the performance of MIMO at the other extreme, i.e., when the available bandwidth is large, which takes us to the regime where the SNR per degree of freedom is low.

†S. Ray (sray@mit.edu), M. Médard (medard@mit.edu) and L. Zheng (lizhong@mit.edu) are with the Laboratory for Information and Decision Systems (LIDS), Massachusetts Institute of Technology, Cambridge, MA 02139. Work is supported by NSF Ultra Wideband Wireless Award ANI0335256, NSF Career Award CCR 0093349 and Hewlett-Packard Award 008542-008.

In wideband channels, the available power is spread over a large number of degrees of freedom. This makes the SNR per degree of freedom low. Hence, while studying these channels, we need to focus on the low SNR regime. We will therefore use the terms “wideband” and “low SNR” interchangeably, with the understanding that the latter refers to the SNR per degree of freedom.

The study of single antenna wideband channels dates back to 1969 and early work has considered the Rayleigh fading channel model. Kennedy [2] shows that the capacity of an infinite bandwidth Rayleigh fading channel is the same as that of an infinite bandwidth additive white Gaussian noise (AWGN) channel with the same average received power. Using the results of Gallager [3], Telatar [4] obtains the capacity per unit energy for the Rayleigh fading channel as a function of bandwidth and signal energy, concluding that given an average power constraint, the Rayleigh fading and AWGN channels have the same capacity in the limit of infinite bandwidth. Telatar and Tse [12] show that this property of the channel capacity is also found in channels with general fading distributions.

Médard and Gallager [6, 15] establish that very large bandwidths yield poor performance for systems that spread the available power uniformly over time and frequency (for example DS-CDMA). They express the input process as an orthonormal expansion of basis functions localized in time and frequency. The energy and fourth moment of the coefficients scale inversely with the bandwidth and square of the bandwidth, respectively. By constraining the fourth moment (as is the case when using spread spectrum signals), they show that mutual information decays to 0 inversely with increasing bandwidth. Telatar and Tse [12] consider a wideband fading channel to be composed of a number of time-varying paths and show that the input signals needed to achieve capacity must be “peaky” in time or frequency. They also show that if white-like signals are used (as for example in spread spectrum communication), the mutual information is inversely proportional to the number of resolvable paths with energy spread out and approaches 0 as the number of paths get large. This does not depend on whether the paths are tracked perfectly at the receiver or not. A strong coding theorem is obtained for this channel in [22]. Subramanian and Hajek [16] derive similar results as [6, 15] using the theory of capacity per unit cost, for a certain fourth-order cost function, called fourthegy.

We now consider the use of multiple antennas over these channels. MIMO channels were first studied from a capacity point of view in [5, 9]. In a Rayleigh flat-fading environment with perfect channel state information (CSI) at the receiver (coherent channel) but no CSI at the transmitter, and statistically independent propagation coefficients between all pairs of transmit and receive antennas, the multiple antenna capacity increases linearly with the smaller of the number of transmit and receive antennas, provided the signal-to-noise ratio is high [5].

When the coherence time of the channel is small (for example if the receiver is mobile),

communication is desirable without training. Here, CSI is unavailable at the transmitter as well as the receiver. This channel is also referred to as the non-coherent channel. In [8], Marzetta and Hochwald derive the structure of the optimal input matrix as a product of two statistically independent matrices; one of them being an isotropically distributed unitary matrix and the other being a diagonal, real and non-negative matrix. They also show that there is no gain, from the point of view of capacity, in having the number of transmit antennas be more than the coherence interval (in symbols) of the channel. Zheng and Tse [14] obtain the non-coherent MIMO capacity in the high SNR regime and show that, in this regime, the number of transmit antennas required need not be more than half the coherence interval (in symbols).

In this paper, we assume that the transmitter and receiver have no channel state information (CSI). Hence, we study the non-coherent channel in this paper. We also assume Rayleigh block fading. In the limit of infinite bandwidth, Zheng and Tse [14] show that the capacities per degree of freedom for the coherent and non-coherent MIMO channels are the same, i.e.,

$$\lim_{\mathsf{SNR}\to 0} \frac{C_{\mathsf{coherent}}(\mathsf{SNR})}{\mathsf{SNR}} = \lim_{\mathsf{SNR}\to 0} \frac{C_{\mathsf{non-coherent}}(\mathsf{SNR})}{\mathsf{SNR}} = r,$$

where, $r$ is the number of receive antennas and $\mathsf{SNR}$ is the average signal-to-noise ratio per degree of freedom at each receive antenna. The capacity can thus be expressed as:

$$C(\mathsf{SNR}) = r\mathsf{SNR} + o(\mathsf{SNR}) \ \text{ nats/channel use}$$

and is thus a linear function only in the limit of low SNR. As SNR increases from 0, capacity increases in a sublinear fashion, showing that low SNR communication is power efficient.

Using a Taylor series expansion, Verdú [17] shows that the second derivative of the capacity at $\mathsf{SNR} = 0$ is finite for the coherent channel. The impact on the coherent capacity of antenna correlation, Ricean factors, polarization diversity and out-of-cell interference is considered in [21]. For the non-coherent channel, Verdú [17] shows that though “flash” signaling is first order optimal, it renders the second derivative $-\infty$. Hence, the coherent and non-coherent channels have the same linear term and differ in their sublinear term. Therefore, the non-coherent channel capacity approaches the wideband limit slower than the coherent channel capacity.

Let us define the sublinear term for the MIMO channel with $t$ transmit and $r$ receive antennas as

$$\Delta^{(t,r)}(\mathsf{SNR}) \triangleq r\mathsf{SNR} - C(\mathsf{SNR}) \ \text{ nats/channel use.}$$

Computing the sublinear term tells us the capacity and also quantifies the convergence of the capacity function to the low SNR limit. Larger the order of the sublinear term, faster the convergence. Using the results of Verdú [17], the sublinear term for the Rayleigh fading coherent MIMO channel, $\Delta^{(t,r)}_{\mathsf{coherent}}(\mathsf{SNR})$, is

$$\Delta^{(t,r)}_{\mathsf{coherent}}(\mathsf{SNR}) = \frac{r(r+t)}{2t}\mathsf{SNR}^2 + o(\mathsf{SNR}^2).$$

On the other extreme, for the i.i.d Rayleigh fading non-coherent MIMO channel, the sublinear term, $\Delta^{(t,r)}_{\mathsf{i.i.d}}(\mathsf{SNR}) \gg O(\mathsf{SNR}^2)$ [17]. In this paper, we compute $\Delta^{(t,r)}_{\mathsf{i.i.d}}(\mathsf{SNR})$ and show that on-off signaling achieves capacity for the i.i.d Rayleigh fading non-coherent MIMO channel. Figure 1 shows the sublinear terms for the Rayleigh fading coherent channel and the i.i.d. Rayleigh fading non-coherent channel with the same number of transmit and receive antennas. A property of the non-coherent capacity is that it tends towards the coherent capacity as the coherence length increases. Hence, the sublinear term for the i.i.d Rayleigh fading non-coherent channel is the largest (non-coherent extreme), whereas, for the coherent channel, it is the smallest (coherent extreme). In this paper, we focus on how the non-coherent MIMO channel capacity is influenced by the coherence length, number of antennas and SNR. We do so, by computing the sublinear term, which in turn tells us the capacity of the low SNR non-coherent MIMO channel of arbitrary coherence length. Thereby, we sweep the region, shown in Figure 1, between the coherent and non-coherent extremes.

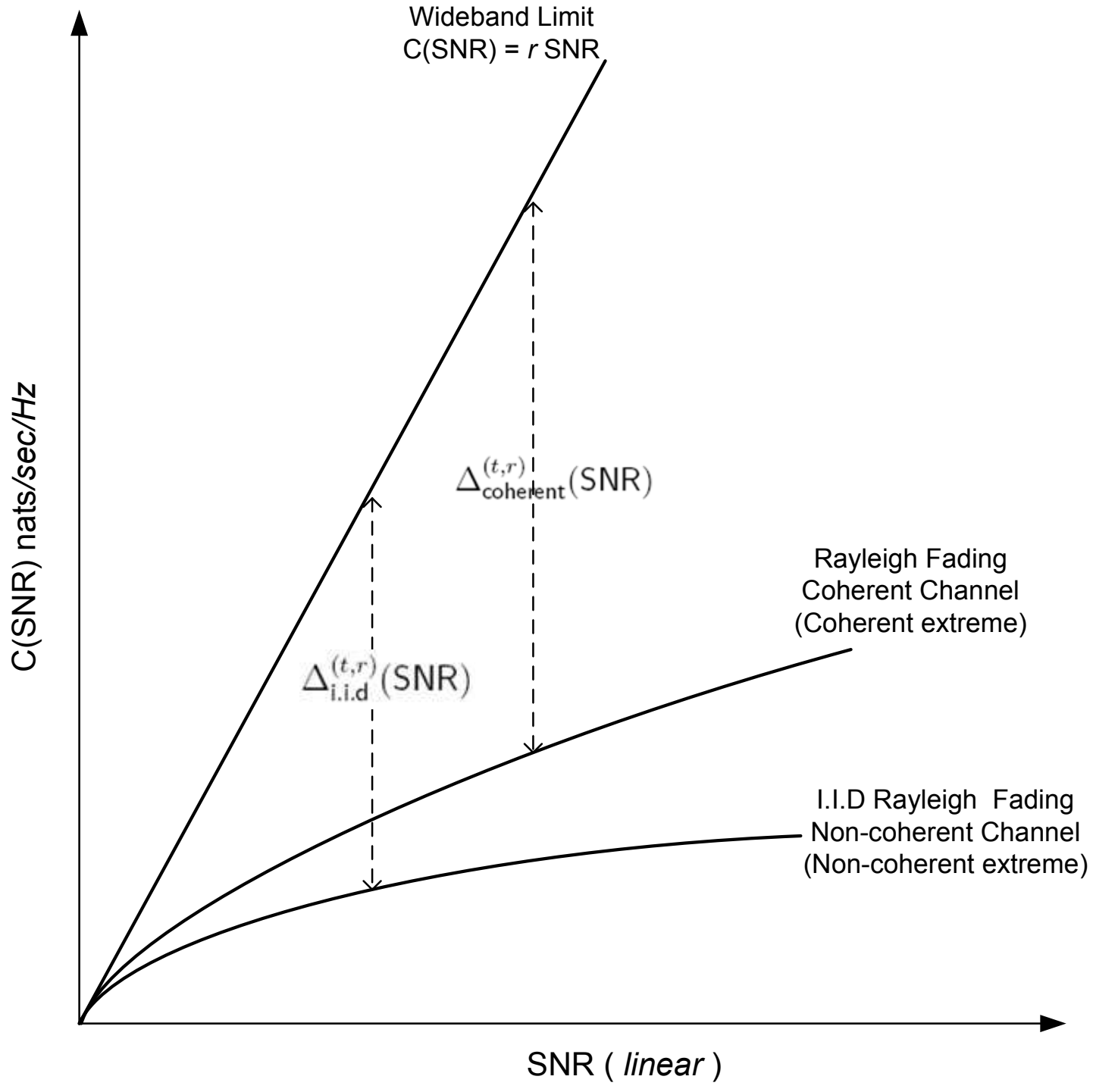

Figure 1: Sublinear capacity term.

In the low SNR regime, the sublinear term also represents the energy efficiency of communication. Let $E_n$ and $N_0$ represent the energy per information nat and the noise spectral level, respectively. We have:

$$\begin{aligned}\frac{E_n}{N_0} &= \frac{\mathsf{SNR}}{C(\mathsf{SNR})} \\ &= \frac{\mathsf{SNR}}{r\mathsf{SNR} - \Delta^{(t,r)}(\mathsf{SNR})} \\ &= \frac{1}{r}\frac{1}{1 - \frac{\Delta^{(t,r)}(\mathsf{SNR})}{r\mathsf{SNR}}}.\end{aligned}$$

Taking logarithms on both sides,

$$\log\left(\frac{E_n}{N_0}\right) \approx \frac{\Delta^{(t,r)}(\mathsf{SNR})}{r\mathsf{SNR}} - \log(r). \tag{1}$$

Equation (1) shows how energy efficiency is related to the sublinear term. The smaller the sublinear term for a channel, the more energy efficient will it be. As the non-coherent capacity is always less than the coherent capacity for the same number of transmit and receive antennas, lack of receiver CSI results in energy inefficiency. Also, note that the minimum energy (in dB) required to reliably transmit one information nat decreases logarithmically with the number of receive antennas.

Let us now turn to Figure 2, which shows how wideband capacity changes with bandwidth. We denote $P$ as the average receive power and $N_0$ as the noise spectral density, which makes the wideband limit $\frac{rP}{N_0}$ nats/sec. We obtain this figure by scaling the y-axis of Figure 1 by the bandwidth. Channels whose capacities converge slowly to the wideband limit have to incur large bandwidth penalties. For the same number of transmit and receive antennas, the non-coherent capacity is less than the coherent capacity. Thus, the non-coherent channel requires a larger bandwidth in order to reliably support the same throughput as the coherent channel. This bandwidth penalty grows with bandwidth. Hence, for the non-coherent channel, the closer we get to the wideband limit, we gain in terms of energy efficiency (as the sublinear term decreases), but the bandwidth penalty becomes larger. We quantify this effect by computing the low SNR non-coherent MIMO capacity. Studying how capacity changes with coherence length also tells us the amount of bandwidth required to achieve a "near coherent" performance. Note that since bandwidth penalty increases with decreasing coherence length, the channel at the non-coherent extreme (i.i.d Rayleigh fading non-coherent channel) has to incur the largest bandwidth penalty.

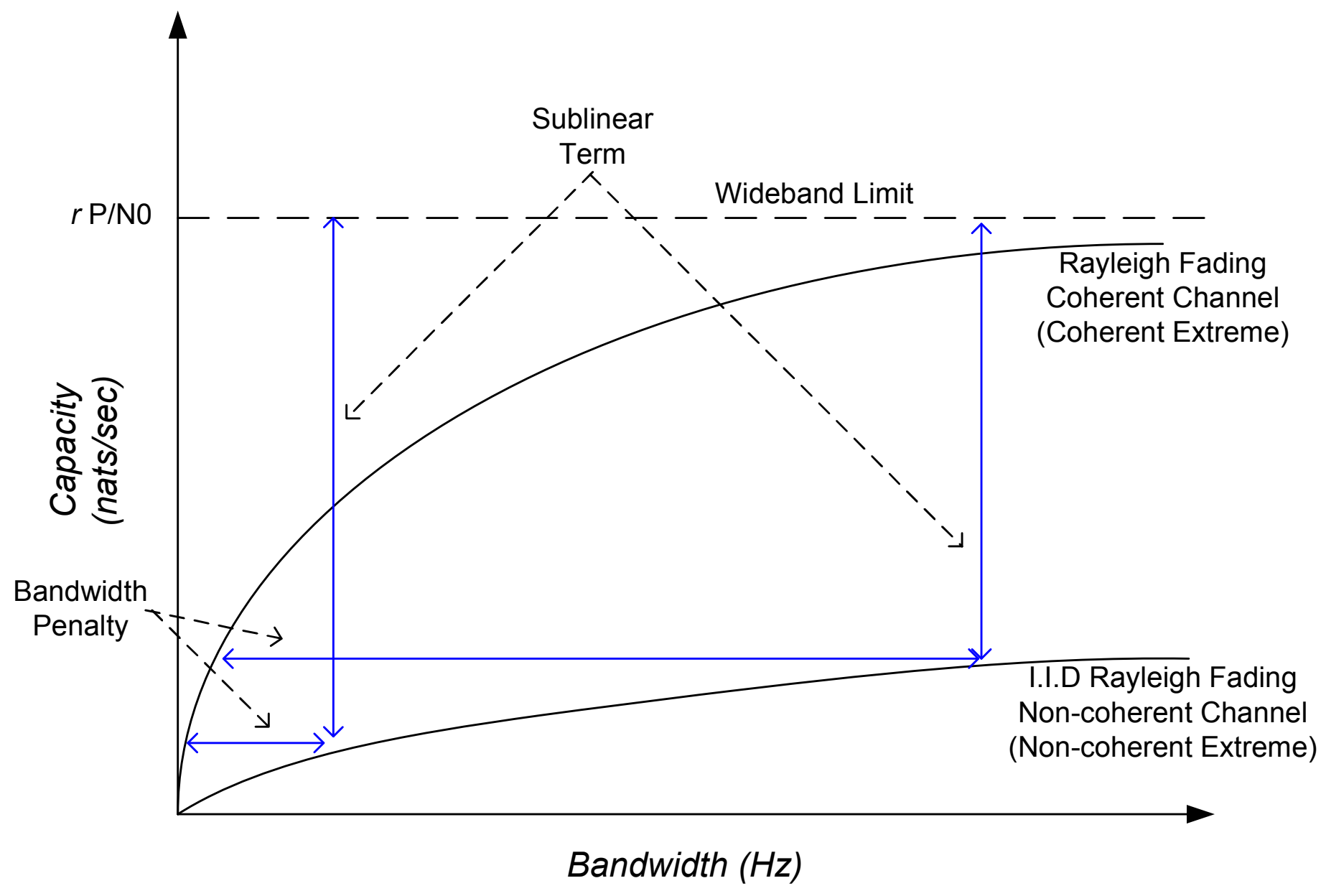

Figure 2: Wideband capacity.

At low SNR, channel estimates become unreliable. Hence, even for slowly varying channels, estimating the channel at the receiver may not be possible. When we have multiple antennas at the receiver as well as the transmitter, estimation becomes much more difficult since there are multiple channel coefficients that need to be estimated. Hence, communication is desirable without training. In [23], non-coherent communication is considered with the input distribution constrained to be exponentially decaying. It is show that the capacity per degree of freedom in the low SNR regime is $O(\mathsf{SNR}^2)$. Reference [24] considers the same capacity under the constraint that only the fourth and sixth order moments of the input are finite. Once again, the non-coherent capacity per degree of freedom is shown to be $O(\mathsf{SNR}^2)$. Hence, [23, 24] show that when there is a higher order (fourth and above) constraint on the input, capacity scales inversely with bandwidth. Thus, the non-coherent capacity does not approach the wideband limit and diverges from the coherent capacity as bandwidth increases. These results are akin to the single antenna channel results [6, 12, 15, 16]. Hassibi and Hochwald [19] propose a training scheme that is near-optimal in the high SNR regime. However, at low SNR, their scheme results in the rate per degree of freedom to go as $O(\mathsf{SNR}^2)$. Since the overall rate decays to 0 inversely with bandwidth, their training scheme is not desirable at low SNR.

In this paper, we consider multiple antenna communication over a wideband, non-coherent Rayleigh block fading channel. We compute the capacity with only an average power constraint, and consider it's interaction with the coherence length of the channel, number of transmit and receive antennas and SNR. We establish how large the coherence length has to be in order for a non-coherent channel to have a "near coherent" performance at low SNR. More specifically, we show that if the channel coherence length is above a certain antenna and SNR dependent threshold, the non-coherent and coherent capacities are the same in the low SNR regime. We show that the transmit antennas effect the sublinear capacity term and hence, the approach of capacity to the wideband limit, with increasing bandwidth. Moreover, we propose a signaling scheme that is near-optimal in the wideband regime.

The capacity problem that we consider in this paper has been considered for single antenna channels by Zheng, Tse and Médard [26]. They consider the interaction between coherence length and capacity at low SNR and compute the order of the sublinear capacity term. The work in this paper builds on their work, where, we analyze the more general MIMO channel and *exactly* compute the sublinear capacity term. We use a finer scale of analysis than [26], which allows us to understand how the transmit and receive antennas effect the sublinear capacity term and hence, the approach of the non-coherent capacity to the wideband capacity limit.

We also analyze the error probability for the non-coherent low SNR MIMO channel. The behavior of error probability for the coherent [7, 20] as well as non-coherent [10, 18] MIMO channels has been well studied in the high SNR regime. For the coherent MIMO channel with coherence length 1, the error exponent is computed by Telatar [9] for any SNR. The behavior of the error exponent for the non-coherent MIMO channel in the low SNR regime has recently been considered by Wu and Srikant in [25]. Their analysis considers the linear capacity term, $r\mathsf{SNR}$, and the error exponent is computed by fixing the coherence length and letting $\mathsf{SNR}$ tend to 0.

Our consideration of the effect of the interaction among SNR, number of transmit and receive antennas and coherence length, on the error probability, yields a more detailed characterization of the error probability behavior than described in [25]. Our analysis shows that in the low SNR regime, the critical rate as well as the cut-off rate are much smaller than the channel capacity. Moreover, the error probability decays inversely with coherence length. We introduce the notion of "diversity" in the low SNR regime and use it to show that error probability decays exponentially with the product of the number of transmit and receive antennas. Hence, in terms of reliability in the wideband regime, transmit antennas have the same importance as receive antennas. In the high SNR regime, it is well known that outage dominates the error probability. Our analysis shows that this is true even at low SNR, i.e., channel outage dominates the error probability at low SNR.

Let us establish notation that will be used in the rest of the paper. The bold type will be used to denote random quantities whereas normal type will be used to denote deterministic ones. Matrices will be denoted by capital letters and the scalar or vector components of matrices will be denoted using appropriate subscripts. Vectors will be represented by small letters with an arrow over them. All vectors are column vectors unless they have a $^T$ superscript. Scalars will be represented by small letters only. The superscript $^\dagger$ will be used to denote the complex conjugate transpose.

The rest of the paper is organized as follows: Section 2 describes the channel model. The capacity and error probability results are in sections 3 and 4, respectively. We conclude in section 5.

# 2 Model

We model the wideband channel as a set of $N$ parallel narrowband channels. In general, the narrowband channels will be correlated. We restrict our analysis in this paper to channels having independent and identical statistics. We also assume that the coherence bandwidth is much larger than the bandwidth of the narrowband channel. Hence, each narrowband channel is modeled as being flat faded. From [26], we see that the behavior of channels with low SNR per degree of freedom is robust to reasonable modeling assumptions and necessary simplifications. Hence, the results for a more precise MIMO channel model will not differ significantly from that of the simple model we consider in this paper.
Using the sampling theorem, the $m^{th}$ narrowband channel at symbol time $k$ can be represented as:

$$\vec{\mathbf{y}}[k,m] = \mathbf{H}[k,m]\vec{\mathbf{x}}[k,m] + \vec{\mathbf{w}}[k,m],$$

where $\mathbf{H}[k,m]$, $\vec{\mathbf{x}}[k,m]$, $\vec{\mathbf{w}}[k,m]$ and $\vec{\mathbf{y}}[k,m]$ are the channel matrix, input vector, noise vector and output vector, respectively, for the $m^{th}$ narrowband channel at symbol time $k$. The pair $(k,m)$ may be considered as an index for the time-frequency slot, or degree of freedom, to communicate. We denote the number of transmit and receive antennas by $t$ and $r$, respectively. Hence, $\vec{\mathbf{x}}[k,m] \in \mathcal{C}^t$ and $\vec{\mathbf{y}}[k,m], \vec{\mathbf{w}}[k,m] \in \mathcal{C}^r$. The channel matrix $\mathbf{H}[k,m]$ is a $r \times t$ complex matrix. The entries of the channel matrix are i.i.d zero-mean complex Gaussian, with independent real and imaginary components. Equivalently, each entry of $\mathbf{H}[k,m]$ has uniformly distributed phase and Rayleigh distributed magnitude. We thus model a Rayleigh fading channel with enough separation within the transmitting and receiving antennas to achieve independence in the entries of $\mathbf{H}[k,m]$. The channel matrix is unknown at the transmitter and the receiver. However, its statistics are known to both. The noise vector $\vec{\mathbf{w}}[k,m]$ is a zero-mean Gaussian vector with the identity as its covariance matrix. Thus, $\vec{\mathbf{w}}[k,m] \backsim \mathcal{CN}(\vec{0}, I_r)$. Since the narrowband channels are assumed to be independent, we will omit the narrowband channel index, $m$, for simplifying notation. The capacity of the wideband channel with power constraint $P$ is thus $N$ times the capacity of each narrowband channel with power constraint $P/N$. Hence, we can focus on the narrowband channel alone. We further assume a block fading channel model, i.e., the channel matrix is random but fixed for the duration of the coherence time of the channel, and is i.i.d across blocks. Hence, we may omit the time index, $k$, and express the narrowband channel within a coherence block of length $l$ symbols as:

$$\mathbf{Y} = \mathbf{H}\mathbf{X} + \mathbf{W},$$

where, $\mathbf{X} \in \mathcal{C}^{t\times l}$ has entries $\mathbf{x}_{ij}, i = 1, ..., t, j = 1, ..., l$, being the signals transmitted from the transmit antenna $i$ at time $j$; $\mathbf{Y} \in \mathcal{C}^{r\times l}$ has entries $\mathbf{y}_{ij}, i = 1, ..., r, j = 1, ..., l$, being the signals received at the receive antenna $i$ at time $j$; the additive noise $\mathbf{W}$ has i.i.d. entries $\mathbf{w}_{ij}$, which are distributed as $\mathcal{CN}(0,1)$. The input $\mathbf{X}$ satisfies the average power constraint

$$\frac{1}{l}E\Big[\mathsf{trace}\Big\{\mathbf{X}\mathbf{X}^\dagger\Big\}\Big] = \mathsf{SNR},$$

where, $\mathsf{SNR}$ is the average signal to noise ratio at each receive antenna per narrowband channel. As $N$ tends to $\infty$, $\mathsf{SNR}$ tends to 0, and the narrowband channel is in the low SNR regime.

# 3 Capacity of the Non-coherent MIMO channel

In this section, we compute the capacity of the non-coherent MIMO channel at low SNR. The analysis shows the interaction between the number of receive and transmit antennas, coherence length of the channel and, SNR in the wideband regime. We also propose a signaling scheme that achieves capacity.

## 3.1 Dependence of capacity on coherence length

We first analyze the dependence of the non-coherent capacity on the coherence length of the channel. In [8], the structure of the capacity achieving input matrix for our non-coherent MIMO channel model is described as

$$\mathbf{X} = \mathbf{A}\mathbf{\Phi},$$

where

$$\mathbf{A} = \begin{bmatrix} \|\vec{\mathbf{x}}_1^T\| & & & & \\ & \cdots & & & \\ & & \|\vec{\mathbf{x}}_i^T\| & & \\ & & & \cdots & \\ & & & & \|\vec{\mathbf{x}}_t^T\| \end{bmatrix},$$

is a $t \times l$ random matrix that is diagonal, real and nonnegative with identically (though may not be independent) distributed entries and $\|\vec{\mathbf{x}}_i^T\|$ is the norm of the signal vector transmitted by the $i^{th}$ antenna. Since these entries are identically distributed, we have $\forall i \in \{1, \ldots, t\}$

$$E[\|\vec{\mathbf{x}}_i^T\|^2] = \frac{l}{t}\mathsf{SNR}.$$

$\mathbf{\Phi}$ is a $l \times l$ isotropically distributed unitary matrix. The row vectors of $\mathbf{\Phi}$ are isotropic random vectors which represent the direction of the signal transmitted from the antennas. $\mathbf{A}$ and $\mathbf{\Phi}$ are statistically independent matrices. Since this structure of the input matrix is optimal, we will restrict our attention to inputs having such structure.
We first prove Lemma 1, which establishes two necessary conditions the input distribution must satisfy for the mutual information of the channel to be above a certain value. This lemma will be used in Theorem 1 to establish the dependence of the non-coherent capacity on the channel coherence length.

**Lemma 1** *For any $\alpha \in (0, 1]$ and $\gamma \in (0, \alpha)$, if there exists an input distribution on $\mathbf{X}$ such that*

$$\frac{1}{l}I(\mathbf{X};\mathbf{Y}) \geq r\mathsf{SNR} - \frac{r(r+t)}{2t}\mathsf{SNR}^{1+\alpha} + O(\mathsf{SNR}^{1+\alpha+\gamma}),$$

*then the following two conditions are satisfied by this distribution:*

$$\frac{t}{l}E\Big[\log(1 + \|\vec{\mathbf{x}}_i^T\|^2)\Big] \;\leq\; \frac{(r+t)}{2t}\mathsf{SNR}^{1+\alpha} + O(\mathsf{SNR}^{1+\alpha+\gamma}), \quad (2)$$

$$tE\left[\log\left(1 + \frac{\|\vec{\mathbf{x}}_i^T\|^2}{l}\right)\right] \;\geq\; \mathsf{SNR} - \frac{(r+t)}{t}\mathsf{SNR}^{1+\alpha} + O(\mathsf{SNR}^{1+\alpha+\gamma}), \quad (3)$$

*for all $i \in \{1, \ldots\ldots, t\}$.*

*Proof:* See Appendix 1. □

We are now ready to prove a theorem that describes the dependence of the non-coherent capacity on the coherence length:

**Theorem 1** *Consider a non-coherent Rayleigh block fading MIMO channel with average signal to noise ratio* $\mathsf{SNR}$*. Let the block length be* $l$ *and the capacity,* $C(\mathsf{SNR})$*. For any* $\alpha \in (0, 1]$ *and* $\gamma \in (0, \alpha)$*, if*

$$C(\mathsf{SNR}) \geq C^*(\mathsf{SNR}) \triangleq r\mathsf{SNR} - \frac{r(r+t)}{2t}\mathsf{SNR}^{1+\alpha} + O(\mathsf{SNR}^{1+\alpha+\gamma}),$$

*then*

$$l > l_{\mathsf{min}} \triangleq \frac{t^2}{(r+t)^2}\mathsf{SNR}^{-2\alpha}.$$

*Proof:* See Appendix 2. □

This theorem states that the coherence length must be strictly larger that $l_{\mathsf{min}}$, for the channel capacity to be above $C^*(\mathsf{SNR})$. Since the inequality for the coherence length is strict, this implies that a channel with capacity $C^*(\mathsf{SNR})$ will have its coherence length, $l^*$, strictly greater than $l_{\mathsf{min}}$, i.e.,

$$l_{\mathsf{min}} < l^*.$$

## 3.2 Communicating using Gaussian-like signals

In this subsection, we propose a signalling scheme using which a rate of $C^*(\mathsf{SNR})$ is achievable if the coherence length is greater than or equal to a threshold, which we denote as $l^G$.
We first prove a lemma that shows that using a Gaussian input distribution, we can achieve "near coherent" performance if the coherence length of the channel is large enough.

**Lemma 2** *Consider a non-coherent Rayleigh block fading MIMO channel with average signal-to-noise ratio* $\mathsf{SNR}$*. Let the block length be* $l$ *and the capacity,* $C(\mathsf{SNR})$*. If we use Gaussian signals over this channel, then for any* $\epsilon \in (0, 1)$*, if*

$$l \geq \frac{t^2}{(r+t)^2}\mathsf{SNR}^{-2(1+\epsilon)},$$

*then*

$$C(\mathsf{SNR}) \geq r\mathsf{SNR} - \frac{r(r+t)}{2t}\mathsf{SNR}^2 + O(\mathsf{SNR}^{2+\epsilon}).$$

*Proof:* We first lower bound the mutual information of the non-coherent channel MIMO channel as

$$\begin{aligned} &I(\mathbf{X};\mathbf{Y}) \\ &\quad = I(\mathbf{X};\mathbf{Y}|\mathbf{H}) + I(\mathbf{H};\mathbf{Y}) - I(\mathbf{H};\mathbf{Y}|\mathbf{X}) \\ &\quad \geq I(\mathbf{X};\mathbf{Y}|\mathbf{H}) - I(\mathbf{H},\mathbf{Y}|\mathbf{X}). \end{aligned} \tag{4}$$

Let us choose the distribution of $\mathbf{X}$ to be one where all the entries of $\mathbf{X}$ are i.i.d. and $\mathcal{CN}(0, \frac{\mathsf{SNR}}{t})$. Note that it is exactly this distribution that achieves capacity for the coherent MIMO channel. Therefore,

$$\frac{1}{l} I(\mathbf{X}; \mathbf{Y}|\mathbf{H}) = r\mathsf{SNR} - \frac{r(r+t)}{2t}\mathsf{SNR}^2 + O(\mathsf{SNR}^3). \tag{5}$$

$I(\mathbf{H}; \mathbf{Y}|\mathbf{X})$ is the information that can be obtained about $\mathbf{H}$ from observing $\mathbf{Y}$, conditioned on $\mathbf{X}$ being known. Therefore

$$\begin{aligned} & I(\mathbf{H}; \mathbf{Y}|\mathbf{X}) \\ & \quad = h(\mathbf{Y}|\mathbf{X}) - h(\mathbf{Y}|\mathbf{X}, \mathbf{H}) \\ & \quad = rtE \log(1 + \|\vec{\mathbf{x}}_i^T\|^2) \\ & \quad \leq rt \log\left(1 + \frac{l}{t}\mathsf{SNR}\right), \end{aligned} \tag{6}$$

where we have used Jensen's inequality to get the upper bound in (6). Combining (4), (5) and (6) and noting that

$$C(\mathsf{SNR}) \geq \frac{1}{l} I(\mathbf{X}; \mathbf{Y}),$$

we obtain:

$$C(\mathsf{SNR}) \geq r\mathsf{SNR} - \frac{r(r+t)}{2t}\mathsf{SNR}^2 - r\frac{t}{l}\log\left(1 + \frac{l}{t}\mathsf{SNR}\right) + O(\mathsf{SNR}^3). \tag{7}$$

For any $\epsilon \in (0, 1]$, let us choose

$$l = \frac{t^2}{(r+t)^2}\mathsf{SNR}^{-2(1+\epsilon)}.$$

Therefore,

$$\begin{aligned} & r\frac{t}{l}\log\left(1 + \frac{l}{t}\mathsf{SNR}\right) \\ & \quad = r\frac{(r+t)^2}{t}\mathsf{SNR}^{2(1+\epsilon)} \log\left(1 + \frac{t}{(r+t)^2\mathsf{SNR}^{1+2\epsilon}}\right) \\ & \quad = r\frac{(r+t)^2}{t}\mathsf{SNR}^{2(1+\epsilon)} \log\left(\frac{t}{(r+t)^2\mathsf{SNR}^{1+2\epsilon}}\right) + o(\mathsf{SNR}^{2(1+\epsilon)}) \\ & \quad = r\frac{(r+t)^2}{t}\mathsf{SNR}^{2(1+\epsilon)} \log\left(\frac{t}{(r+t)^2}\right) + r\frac{(r+t)^2}{t}(1+2\epsilon)\mathsf{SNR}^{2+\epsilon}\left[\mathsf{SNR}^{\epsilon}\log\left(\frac{1}{\mathsf{SNR}}\right)\right] + o(\mathsf{SNR}^{2(1+\epsilon)}) \\ & \quad \leq r\frac{(r+t)^2}{t}\mathsf{SNR}^{2(1+\epsilon)} \log\left(\frac{t}{(r+t)^2}\right) + r\frac{(r+t)^2}{t}(1+2\epsilon)\mathsf{SNR}^{2+\epsilon} + o(\mathsf{SNR}^{2(1+\epsilon)}) \qquad (8) \\ & \quad \equiv O(\mathsf{SNR}^{2+\epsilon}). \end{aligned}$$

In (8), we use that, since $\epsilon > 0$ and $\mathsf{SNR} \to 0$, $\mathsf{SNR}^{\epsilon} \log(\frac{1}{\mathsf{SNR}}) \ll 1$. Since $r\frac{t}{l}\log\left(1+\frac{l}{t}\mathsf{SNR}\right)$ decreases monotonically with $l$, we have that

$$l \geq \frac{t^2}{(r+t)^2}\mathsf{SNR}^{-2(1+\epsilon)}$$
$$\Rightarrow r\frac{t}{l}\log\left(1+\frac{l}{t}\mathsf{SNR}\right) \leq O(\mathsf{SNR}^{2+\epsilon}).$$

Combining this with (7) completes the proof. □

We now introduce an input distribution that has a flashy as well as a continuous nature. A similar input distribution was first introduced in [26] for achieving the *order* of the sublinear capacity term for a single-input, single-output, non-coherent Rayleigh block fading channel. For a given $\alpha \in (0,1]$, let us transmit in only $\delta(\mathsf{SNR}) = \mathsf{SNR}^{1-\alpha}$ fraction of the blocks. As we are in the low signal to noise ratio regime, $\delta(\mathsf{SNR}) \in (\mathsf{SNR}, 1]$. Since we concentrate the power only over a fraction of the blocks, the signal to noise ratio for the blocks in which we transmit increases to $\mathsf{SNR}^{'}$ where

$$\mathsf{SNR}^{'} \triangleq \frac{\mathsf{SNR}}{\delta(\mathsf{SNR})} = \mathsf{SNR}^{\alpha}.$$

In the blocks that we choose to transmit, let the entries of the input matrix $\mathbf{X}$ be i.i.d. $\mathcal{CN}(0, \frac{\mathsf{SNR}^{'}}{t})$.
Note that as we increase $\alpha$ from 0 to 1, the fraction of blocks that we transmit increases from $\mathsf{SNR}$ to 1. Therefore, as $\alpha$ increases, the distribution changes from a peaky to a continuous one. We will call this type of signalling as *Peaky Gaussian*. We prove the following theorem:

**Theorem 2** *Consider a non-coherent Rayleigh block fading MIMO channel with average signal to noise ratio* $\mathsf{SNR}$. *Let the block length be $l$ and the capacity, $C(\mathsf{SNR})$. If we use Peaky Gaussian signals over this channel, then for any $\alpha \in (0,1]$ and $\epsilon \in (0,\alpha)$, if*

$$l \geq l^G \triangleq \frac{t^2}{(r+t)^2}\mathsf{SNR}^{-2(\alpha+\epsilon)},$$

*then*

$$C(\mathsf{SNR}) \geq C^*(\mathsf{SNR}) = r\mathsf{SNR} - \frac{r(r+t)}{2t}\mathsf{SNR}^{1+\alpha} + O(\mathsf{SNR}^{1+\alpha+\epsilon}).$$

*Proof:* Let us use the Peaky Gaussian like distribution for communicating over the non-coherent MIMO channel. We can now apply Lemma 2 to the blocks that we choose to transmit. Note that these blocks have a signal to noise ratio of $\mathsf{SNR}^{'}$. Thus, for any $\epsilon^{'} \in (0,1]$, if

$$l \geq \frac{t^2}{(r+t)^2}\mathsf{SNR}^{'-2(1+\epsilon^{'})}$$
$$= \frac{t^2}{(r+t)^2}\mathsf{SNR}^{-2(\alpha+\alpha\epsilon^{'})}.$$

then

$$C(\mathsf{SNR}^{'}) \geq r\mathsf{SNR}^{'} - \frac{r(r+t)}{2t}\mathsf{SNR}^{'2} + O(\mathsf{SNR}^{'2+\epsilon^{'}}).$$

Since we are transmitting in $\delta(\mathsf{SNR})$ fraction of the blocks,

$$\begin{aligned} &C(\mathsf{SNR}) \\ &\quad = \delta(\mathsf{SNR})C(\mathsf{SNR}^{'}) \\ &\quad \geq r\mathsf{SNR} - \frac{r(r+t)}{2t}\mathsf{SNR}^{1+\alpha} + O(\mathsf{SNR}^{1+\alpha+\alpha\epsilon^{'}}). \end{aligned}$$

Note that for $\epsilon^{'} \in (0,1]$, $\alpha\epsilon^{'} \triangleq \epsilon \in (0,\alpha]$. This completes the proof. □

Thus, we see that using Peaky Gaussian signals, a rate of $C^{*}(\mathsf{SNR})$ is achievable if the coherence length is greater than or equal to $l^{G}$.
To reliably achieve any rate, the required coherence length using Peaky Gaussian signaling is strictly greater than the required length (Theorem 1) using the optimal input distribution. Thus, if $l^{*}$ is the coherence length needed to have a capacity of $C^{*}(\mathsf{SNR})$,

$$l_{\mathsf{min}} < l^{*} \leq l^{G}.$$

However for $\alpha \in (0,1]$, as $\epsilon \to 0$, $l^{G} \to l_{\mathsf{min}}$. Hence, the Peaky Gaussian input distribution is near-optimal for the non-coherent MIMO channel.
Thus, from Theorems 1 and 2, we see that for any $\alpha \in (0,1]$ and $\epsilon \in (0,\alpha)$, if

$$\frac{t^2}{(r+t)^2}\mathsf{SNR}^{-2\alpha} < l \leq \frac{t^2}{(r+t)^2}\mathsf{SNR}^{-2(\alpha+\epsilon)}, \tag{9}$$

the sublinear capacity term is:

$$\Delta^{(t,r)}(\mathsf{SNR}) = \frac{r(r+t)}{2t}\mathsf{SNR}^{1+\alpha} + O(\mathsf{SNR}^{1+\alpha+\epsilon}).$$

We summarize this result in the following theorem:

**Theorem 3** *Consider a non-coherent Rayleigh block fading MIMO channel with average signal to noise ratio* $\mathsf{SNR}$*. For any* $\alpha \in (0,1]$ *and* $\epsilon \in (0,\alpha)$*, the capacity of the channel is*

$$C(\mathsf{SNR}) = r\mathsf{SNR} - \frac{r(r+t)}{2t}\mathsf{SNR}^{1+\alpha} + O(\mathsf{SNR}^{1+\alpha+\epsilon})$$

*if and only if there exists a* $\sigma \in (0,\epsilon)$ *such that*

$$l = \frac{t^2}{(r+t)^2}\mathsf{SNR}^{-2(\alpha+\sigma)}.$$

This theorem tells us the capacity of a non-coherent MIMO channel in the low SNR regime and shows its dependence on the coherence length of the channel, number of receive and transmit antennas and SNR. Note that the transmit antennas effect the sublinear capacity term. Peaky Gaussian signals are near-optimal when communicating over this channel. Note that $\sigma$ is used in the theorem to parameterize (9). The theorem leads to the following corollary:

**Corollary 1** *Consider a non-coherent Rayleigh block fading MIMO channel with average signal to noise ratio* $\mathsf{SNR}$*. For any* $\alpha \in (0, 1]$ *and* $\epsilon \in (0, \alpha)$*, the sublinear capacity term*

$$\Delta^{(t,r)}(\mathsf{SNR}) = \frac{r(r+t)}{2t}\mathsf{SNR}^{1+\alpha} + O(\mathsf{SNR}^{1+\alpha+\epsilon})$$

*if and only if there exists a* $\sigma \in (0, \epsilon)$ *such that*

$$l = \frac{t^2}{(r+t)^2}\mathsf{SNR}^{-2(\alpha+\sigma)}.$$

In Theorem 3 and Corollary 1, $\alpha$ is used to indicate how close the channel capacity is to the coherent and non-coherent extremes. The coherent channel corresponds to the case when $\alpha = 1$ and the i.i.d non-coherent channel corresponds to the case when $\alpha \to 0$. We have also seen that Peaky Gaussian signals are optimal for the non-coherent MIMO channel. Thus, with a channel coherence length of $l \sim \frac{t^2}{(r+t)^2}\mathsf{SNR}^{-2\alpha}$, one should transmit Gaussian signals in $\delta = \mathsf{SNR}^{1-\alpha}$ fraction of the blocks. At the coherent extreme, $\delta = \mathsf{SNR}^0$ and one should transmit in all the blocks in order to achieve capacity. On the other hand, for the i.i.d Rayleigh fading channel (non-coherent extreme), one should only transmit in $\delta = \mathsf{SNR}^1$ fraction of the blocks. We shall study the non-coherent extreme with a finer scaling later on in the paper.

Let us eliminate the parameter $\alpha$ from Corollary 1. Hence, the sublinear capacity term becomes

$$\Delta^{(t,r)}(\mathsf{SNR}) = \frac{r}{2\sqrt{l}}\mathsf{SNR} + o(\frac{\mathsf{SNR}}{\sqrt{l}}).$$

From (1), we have

$$\log\left(\frac{E_n}{N_0}\right) \propto \sqrt{\frac{1}{l}}.$$

Hence, the minimum energy required to transmit an information bit decreases inversely with the square root of the coherence length of the channel. Thus, energy efficiency improves as the coherence length increases. These results apply only for $\alpha \in (0, 1]$. For channels whose coherence time is larger that $\frac{t^2}{(t+r)^2}\mathsf{SNR}^{-2}$, the sublinear capacity term remains $O(\mathsf{SNR}^2)$. We now focus on the coherent and non-coherent extremes.

## 3.3 Coherent Extreme

In this case, $\alpha = 1$ and from Theorem 3, we know that for $\epsilon \in (0, 1)$

$$C(\mathsf{SNR}) = r\mathsf{SNR} - \frac{r(r+t)}{2t}\mathsf{SNR}^2 + O(\mathsf{SNR}^{2+\epsilon})$$

iff there exists a $\sigma \in (0, \epsilon)$ such that

$$l = \frac{t^2}{(r+t)^2}\mathsf{SNR}^{-2(1+\sigma)}. \tag{10}$$

We see that provided the coherence length is large enough, the non-coherent capacity is the same as the coherent capacity in the low SNR regime. Moreover, the Peaky Gaussian signal is now completely continuous. Hence, when $l \geq \frac{t^2}{(r+t)^2}\mathsf{SNR}^{-2}$, the coherent and non-coherent capacities are the same in the low SNR regime and, continuous Gaussian signals are optimal for both.

### 3.4 Non-coherent Extreme

From Theorem 3, we see that as $\alpha \to 0$, $l \to 1$ and we have an i.i.d. Rayleigh fading channel. In order to get the exact value of the sublinear capacity term for this channel, we need to know the precise value of $\alpha$, which is not possible by this asymptotic analysis. We do the precise analysis in Appendix 3 and show that the capacity is[1]

$$C(\mathsf{SNR}) = r\mathsf{SNR} - \Delta^{(t,r)}_{\mathsf{i.i.d}}(\mathsf{SNR}).$$

where,

$$\Delta^{(t,r)}_{\mathsf{i.i.d}}(\mathsf{SNR}) \doteq \frac{r\mathsf{SNR}}{\log(\frac{r}{\mathsf{SNR}})}.$$

The capacity is achieved using an on-off input distribution that becomes increasingly "flashy" at low SNR. This is consistent with our asymptotic analysis which showed that only $\delta = \mathsf{SNR}^1$ fraction of the blocks should be used for transmission in the non-coherent extreme. Hence, the result shows that besides on-off signaling being optimal for the single-input, single-output i.i.d Rayleigh fading channel [13], it is also capacity achieving when multiple antennas are used.

## 4 Error probability for the non-coherent MIMO channel

In this section, we study the block error probability for the non-coherent MIMO channel, $P^{\mathsf{block}}_{\mathsf{error}}$, when maximum-likelihood decoding is used at the receiver. This error probability is the average over the ensemble of codes when Peaky Gaussian signaling is used and can be expressed as:

$$\begin{aligned} P^{\mathsf{block}}_{\mathsf{error}} &= \Pr(\text{error}|\text{Block used for transmission}) \cdot \Pr(\text{Block used for transmission}) \\ &\quad + \Pr(\text{error}|\text{Block not used for transmission}) \cdot \Pr(\text{Block not used for transmission}). \end{aligned}$$

[1]Definition of ($\doteq$): Let $f(\mathsf{SNR})$ and $g(\mathsf{SNR})$ be functions of $\mathsf{SNR}$. We denote $f(\mathsf{SNR}) \doteq g(\mathsf{SNR})$ if

$$\lim_{\mathsf{SNR}\to 0} \frac{\log f(\mathsf{SNR})}{\log g(\mathsf{SNR})} = 1.$$

Since we use Peaky Gaussian signaling and the receiver is assumed to have perfect knowledge of the blocks that are being used for transmission, we have

$$\begin{aligned} \Pr(\text{Block used for transmission}) &= \delta(\mathsf{SNR}), \\ \Pr(\text{error}|\text{Block not used for transmission}) &= 0. \end{aligned}$$

Hence,

$$P_{\mathsf{error}}^{\mathsf{block}} = \delta(\mathsf{SNR}) \cdot \Pr(\text{error}|\text{Block used for transmission}).$$

If we consider the input matrix transmitted in a block, $\mathbf{X}$, as a super symbol of dimension $t \times l$, the channel is memoryless, since, for each use of the channel an independent realization of $\mathbf{H}$ is drawn. Hence, using the results in [1], the error probability can be bounded as

$$\Pr(\text{error}|\text{Block used for transmission}) \leq \exp[-E_r(R)],$$

where, $E_r(R)$ is the random coding error exponent for the super symbol channel:

$$E_r(R) = \max_{\rho \in [0,1]} \Big\{ E_0(\rho) - \rho R \Big\},$$

where,

$q(X)$ is the distribution of $\mathbf{X}$, $R$ is the transmission rate in nats per block used for transmission and $\mathbf{Y}$ is the channel's output matrix.
Since the signaling is Gaussian in the block used for transmission,

$$q(X) = \frac{1}{\pi^{lt}} \exp\Big[ - \mathsf{trace}(X^\dagger X) \Big].$$

The range of $R$ for which $E_r(R)$ is positive is:

$$0 \leq R \leq \frac{l}{\delta(\mathsf{SNR})} \cdot C(\mathsf{SNR}) \triangleq C^{\mathsf{block}}(\mathsf{SNR}), \tag{11}$$

where, $C^{\mathsf{block}}(\mathsf{SNR})$ is the non-coherent capacity per block.
If we express $l$ as

$$l = \frac{t^2}{(r+t)^2} \mathsf{SNR}^{-2\nu}, \quad \nu > 0,$$

then, from the results in the capacity section,

$$\begin{aligned} \delta(\mathsf{SNR}) &= \mathsf{SNR}^{1-\min\{1,\nu\}}, \\ C(\mathsf{SNR}) &= r\mathsf{SNR} - \frac{r(r+t)}{2t}\mathsf{SNR}^{1+\min\{1,\nu\}} + o(\mathsf{SNR}^{1+\min\{1,\nu\}}), \\ C^{\mathsf{block}}(\mathsf{SNR}) &= \frac{t^2}{(r+t)^2}\mathsf{SNR}^{-2\nu} \left[ r\mathsf{SNR}^{\min\{1,\nu\}} - \frac{r(r+t)}{2t}\mathsf{SNR}^{2\min\{1,\nu\}} \right] + o(\mathsf{SNR}^{-2[\nu - \min\{1,\nu\}]}). \end{aligned}$$

The signal-to-noise ratio in the block used for transmission $\mathsf{SNR}_b$, is:

$$\mathsf{SNR}_b \triangleq \frac{\mathsf{SNR}}{\delta(\mathsf{SNR})} = \mathsf{SNR}^{\min\{1,\nu\}}.$$

The main result is summarized in the following theorem, which is proved in subsection 4.2:

**Theorem 4** *The block error probability for a non-coherent Rayleigh block fading MIMO channel, $P_{\mathsf{error}}^{\mathsf{block}}$, when maximum-likelihood decoding is used at the receiver can be upper bounded as:*

$$P_{\mathsf{error}}^{\mathsf{block}} \leq [\mathsf{SNR}^{1-\min\{1,\nu\}}] \cdot \exp[-E_r(R)],$$

*where,*

$$\begin{aligned} E_r(R) &= rt\log\left(1+\frac{t\mathsf{SNR}^{-[2\nu-\min\{1,\nu\}]}}{2(t+r)^2}\right) - R - o(1) && R \in [0, R_{\mathsf{critical}}] \\ &= rt\log\left(1+\frac{\rho^* t\mathsf{SNR}^{-[2\nu-\min\{1,\nu\}]}}{(t+r)^2(1+\rho^*)}\right) - \rho^* R - o(1) && R \in [R_{\mathsf{critical}}, C_{\mathsf{T,lb}}^{\mathsf{block}}(\mathsf{SNR})] \\ &= o(1) && R \in [C_{\mathsf{T,lb}}^{\mathsf{block}}(\mathsf{SNR}), C^{\mathsf{block}}(\mathsf{SNR})] \\ &= 0 && R \in [C^{\mathsf{block}}(\mathsf{SNR}), \infty), \end{aligned}$$

*and*

$$\begin{aligned} \rho^* &= \frac{1}{2}\left[\sqrt{1+4\left(\frac{rt}{R} - \frac{(t+r)^2\mathsf{SNR}^{2\nu-\min\{1,\nu\}}}{t}\right)} - 1\right], \\ R_{\mathsf{critical}} &= rt/2 + o(1), \\ C_{\mathsf{T,lb}}^{\mathsf{block}}(\mathsf{SNR}) &= \frac{t^2}{(r+t)^2}\mathsf{SNR}^{-2\nu}\Big[r\mathsf{SNR}^{\min\{1,\nu\}} - 2\frac{r(r+t)}{\sqrt{t}}\mathsf{SNR}^{\nu+\frac{\min\{1,\nu\}}{2}} \\ &\quad -\frac{r(r+t)}{2t}\mathsf{SNR}^{2\min\{1,\nu\}} + o\Big(\mathsf{SNR}^{\min\{\nu+\frac{\min\{1,\nu\}}{2}, 2\min\{1,\nu\}\}}\Big)\Big]. \end{aligned}$$

## 4.1 Discussion of Theorem 4

Theorem 4 divides the range of rates for which $E_r(R)$ is positive into three regions - A, B and C, which is illustrated in Figure 3. Let us consider region A: $R \in [0, R_{\mathsf{critical}}]$. Since, $R_{\mathsf{critical}} = O(1)$ and $C^{\mathsf{block}}(\mathsf{SNR}) = O(\mathsf{SNR}^{-[2\nu-\min\{1,\nu\}]})$, the critical rate is much smaller than the channel capacity:

$$R_{\mathsf{critical}} \ll C^{\mathsf{block}}(\mathsf{SNR}).$$

Region A is an $O(\mathsf{SNR}^{[2\nu-\min\{1,\nu\}]})$ fraction of the capacity and is very small in the wideband regime. The cut-off rate, $R_{\mathsf{cut-off}}$, is given by

$$R_{\mathsf{cut-off}} = E_r(0) \doteq rt \cdot [2\nu - \min\{1,\nu\}] \cdot \log\left(\frac{1}{\mathsf{SNR}}\right).$$

Since the cut-off rate is an $O\Big(\mathsf{SNR}^{[2\nu-\min\{1,\nu\}]} \cdot \log(\frac{1}{\mathsf{SNR}})\Big)$ fraction of the capacity, it is much smaller than the capacity in the wideband regime:

$$R_{\mathsf{cut-off}} \ll C^{\mathsf{block}}(\mathsf{SNR}).$$

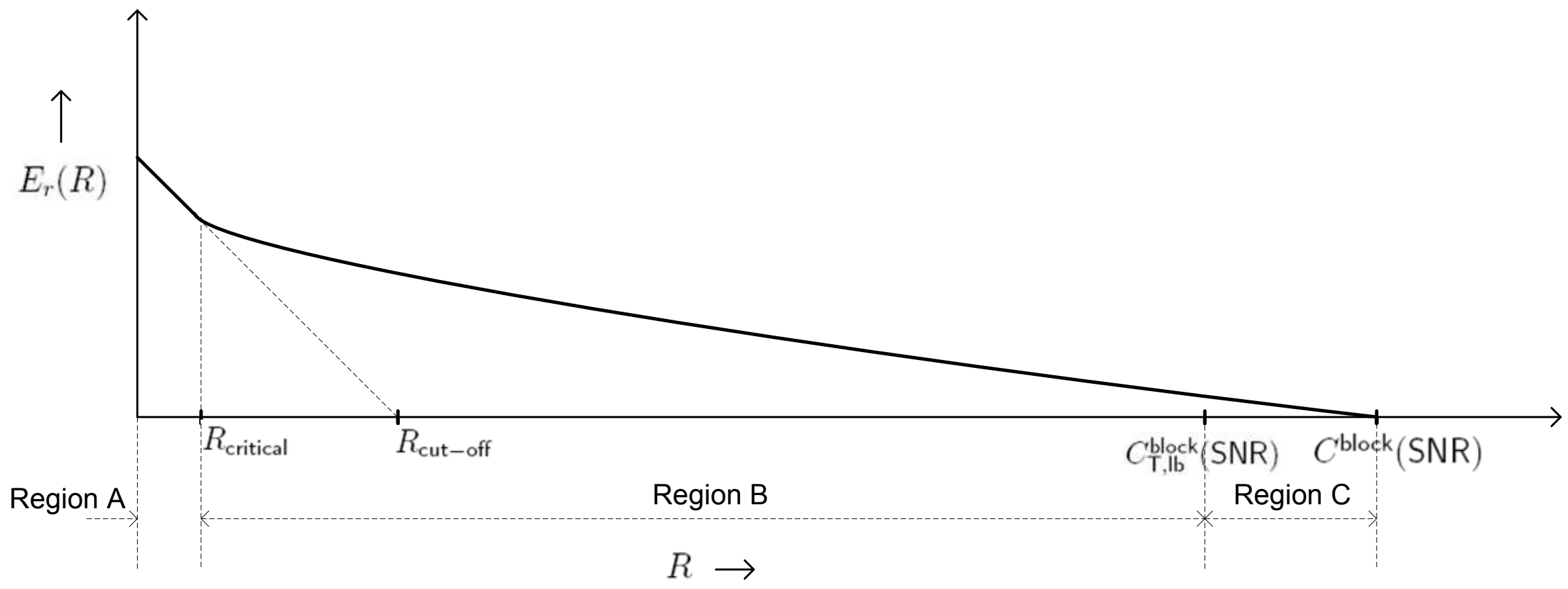

Figure 3: Random coding error exponent for the non-coherent MIMO channel at low SNR.

Let us consider the third region over which $E_r(R)$ is positive, region C: $R \in [C^{\mathsf{block}}_{\mathsf{T,lb}}(\mathsf{SNR}), C^{\mathsf{block}}(\mathsf{SNR})]$. This interval is a $[C^{\mathsf{block}}(\mathsf{SNR}) - C^{\mathsf{block}}_{\mathsf{T,lb}}(\mathsf{SNR})]/C^{\mathsf{block}}(\mathsf{SNR})$ fraction of the capacity where,

$$\frac{C^{\mathsf{block}}(\mathsf{SNR}) - C^{\mathsf{block}}_{\mathsf{T,lb}}(\mathsf{SNR})}{C^{\mathsf{block}}(\mathsf{SNR})} = \begin{cases} O(\mathsf{SNR}^{\nu - \frac{\min\{1,\nu\}}{2}}) & \nu \le \frac{3}{2} \\ o(\mathsf{SNR}) & \nu > \frac{3}{2} \end{cases}$$

Hence, region C is also a very small fraction of the capacity in the wideband regime. Therefore, we can conclude that it is region B: $R \in [R_{\mathsf{critical}}, C^{\mathsf{block}}_{\mathsf{T,lb}}(\mathsf{SNR})]$, that dominates the range of rates in the wideband regime.
From Theorem 4, the error probability in Region B can be expressed as:

$$P^{\mathsf{block}}_{\mathsf{error}} \sim \mathsf{SNR}^{1-\min\{1,\nu\}} \cdot \left[\frac{R}{l \cdot \mathsf{SNR}^{\min\{1,\nu\}}}\right]^{rt}. \tag{12}$$

To observe this, let us consider the error exponent for

$$R = l \cdot r\mathsf{SNR}^{\kappa}, \quad \min\{1,\nu\} < \kappa < 2\nu. \tag{13}$$

This rate lies in Region B and the optimum $\rho$ is

$$\rho^* = O\Big(\frac{1}{R}\Big).$$

Substituting in Theorem 4, we observe

$$\begin{aligned} &E_r(R) \\ &= rt \log\left(1 + \frac{\rho^* t \mathsf{SNR}^{-[2\nu - \min\{1,\nu\}]}}{(t+r)^2(1+\rho^*)}\right) - \rho^* R - o(1)\Big|_{(\rho^* = O(\frac{1}{R}))} \\ &\doteq rt \log\left[\frac{l \cdot \mathsf{SNR}^{\min\{1,\nu\}}}{R}\right]. \end{aligned}$$

For $\nu \leq 1$, $\mathsf{SNR}^{\min\{1,\nu\}} \propto 1/\sqrt{l}$. Hence, for a fixed rate $R$, the error probability decays *inversely* with the coherence length in the following way:

$$P_{\mathsf{error}}^{\mathsf{block}} \propto \begin{cases} \left(\frac{1}{l}\right)^{\frac{rt-1}{2}} & l \leq \mathsf{SNR}^{-2} \\ \left(\frac{1}{l}\right)^{rt} & l > \mathsf{SNR}^{-2} \end{cases}$$

Let us now examine the effect of antennas on the error probability. To analyze this, we propose a definition of "diversity" in the low SNR / wideband regime.
Let $\mathcal{P}$ and $W$ be the total received power and system bandwidth, respectively. High SNR diversity, $d_H(W)$, is commonly defined as:

$$d_H(W) \triangleq -\lim_{\mathcal{P}\to\infty} \frac{\log(P_{\mathsf{error}}^{\mathsf{block}})}{\log(\mathcal{P})}.$$

This definition describes the asymptotic behavior of error probability with received power, for fixed bandwidth.
In the low SNR/wideband regime, we define diversity, $d_L(\mathcal{P})$, as:

$$d_L(\mathcal{P}) \triangleq -\lim_{W\to\infty} \frac{\log(P_{\mathsf{error}}^{\mathsf{block}})}{\log(W)}.$$

This definition describes the asymptotic behavior of error probability with bandwidth, for fixed received power. Since, $\mathsf{SNR} \propto 1/W$, an equivalent definition of low SNR diversity is [2]:

$$d_L \triangleq \lim_{\mathsf{SNR}\to 0} \frac{\log(P_{\mathsf{error}}^{\mathsf{block}})}{\log(\mathsf{SNR})}. \tag{14}$$

From (12, 13), we have

$$d_L = r \cdot t \cdot \Big[\kappa - \min\{1,\nu\}\Big] + 1 - \min\{1,\nu\}.$$

Hence, we conclude that the decay in error probability is *exponential* with the product of the number of transmit and receive antennas, $r \cdot t$. Similar to the high SNR regime, the product of the number of transmit and receive antennas comes about as a diversity factor in the low SNR regime. Hence, we conjecture that $r \cdot t$ is a diversity factor for a MIMO channel at any SNR.
In the capacity section of this paper, we have seen that receive antennas have greater significance than transmit antennas since, the former effects the linear as well as the sublinear capacity term whereas, the latter effects only the sublinear term. However, since the error probability decays exponentially with $r \cdot t$, the transmit antennas have the same importance as receive antennas in terms of reliability. This emphasizes the importance of multiple transmit antennas in the wideband regime.
Let us now consider channel outage in the low SNR regime. For a block fading channel,

[2]We omit the argument of $d_L(.)$ for simplicity.

outage occurs in a coherence block when the channel matrix is so ill-conditioned that the block mutual information cannot support the target block data rate. We denote the outage probability as $P_{\mathsf{outage}}$ and present a heuristic computation to show that

$$P_{\mathsf{error}}^{\mathsf{block}} \sim \mathsf{SNR}^{1-\min\{1,\nu\}} \cdot P_{\mathsf{outage}}.$$

Thus, we see that in the low SNR/wideband regime, for rates away from capacity, the error probability is dominated by the outage probability. Hence, like at high SNR, channel outage is the major source for errors even at low SNR.
*Heuristic Proof:* The outage probability can be upper bounded using a training based scheme (This scheme is described in detail in the proof of Theorem 4). We directly state the channel model when this scheme is used for data transmission (the first $t$ symbols are used for training):

$$\vec{\mathbf{y}}_i = \mathbf{H}^{'}\vec{\mathbf{x}}_i + \vec{\mathbf{v}}_i^{'}, \quad i = t+1, \ldots, l,$$

where, $\mathbf{H}^{'}$ has i.i.d $\mathcal{C}N(0,1)$ entries and is perfectly known at the receiver (this is the MMSE channel estimate) , $\vec{\mathbf{v}}_i^{'}$ is a zero-mean noise vector having the covariance matrix

$$E[\vec{\mathbf{v}}_i^{'}\vec{\mathbf{v}}_i^{'\dagger}] = I_r,$$

and, $\{\vec{\mathbf{x}}_i\}$ are i.i.d complex Gaussian vectors:

$$\vec{\mathbf{x}}_i \sim \mathcal{CN}(0, \frac{f^*(\mathsf{SNR})}{t} \cdot I_t),$$

where,

$$f^*(\mathsf{SNR}) = \mathsf{SNR}^{\min\{1,\nu\}} + o(\mathsf{SNR}^{\min\{1,\nu\}}).$$

Now,

$$\begin{aligned}
&P_{\mathsf{outage}} \\
&\quad = \Pr\left(I(\vec{\mathbf{x}}_{t+1}, \ldots, \vec{\mathbf{x}}_l; \vec{\mathbf{y}}_{t+1}, \ldots, \vec{\mathbf{y}}_l | \mathbf{H}^{'}) < R\right) \\
&\quad \leq \Pr\left(\log\det\left(I_t + \frac{f^*(\mathsf{SNR})}{t}\mathbf{H}^{'\dagger}\mathbf{H}^{'}\right) < \frac{R}{l-t}\right) && (15) \\
&\quad \leq \Pr\left(\log\left(1 + \frac{f^*(\mathsf{SNR})}{t}\mathsf{trace}(\mathbf{H}^{'\dagger}\mathbf{H}^{'})\right) < \frac{R}{l-t}\right) && (16) \\
&\quad \sim \Pr\left(\chi^2_{rt} < \frac{R}{lf^*(\mathsf{SNR})}\right). && (17)
\end{aligned}$$

Equation (15) follows since the mutual information is minimized if $\{\vec{\mathbf{v}}_i^{'}\}$ are i.i.d complex Gaussian [11, 19]. In (16), we use the inequality:

$$\det\left(I_t + \frac{f^*(\mathsf{SNR})}{t}\mathbf{H}^{'\dagger}\mathbf{H}^{'}\right) \geq 1 + \frac{f^*(\mathsf{SNR})}{t}\mathsf{trace}(\mathbf{H}^{'\dagger}\mathbf{H}^{'}).$$

In (17), $\chi^2_{rt}$ represents $\mathsf{trace}(\mathbf{H}'^{\dagger}\mathbf{H}')$ and is a chi-squared random variable with $rt$ degrees of freedom. Hence, if we choose the rate in Region B as in (13), we have for low SNR,

$$\frac{R}{lf^*(\mathsf{SNR})} \ll 1.$$

Hence,

$$\begin{aligned} P_{\mathsf{outage}} &\sim \left[\frac{R}{l\cdot\mathsf{SNR}^{\min\{1,\nu\}}}\right]^{rt} \\ \Rightarrow P^{\mathsf{block}}_{\mathsf{error}} &\sim \mathsf{SNR}^{1-\min\{1,\nu\}}\cdot P_{\mathsf{outage}}. \end{aligned}$$

□

## 4.2 Proof of Theorem 4

**Upper Bound to $E_r(R)$:**
We first establish an upper bound to $E_r(R)$ by providing the receiver perfect knowledge of $\mathbf{H}$. Let us denote the random coding error exponent for this coherent channel by $E^U_r(R)$. Since, the error probability for the coherent channel cannot be greater that the channel without knowledge of $\mathbf{H}$, we have

$$E_r(R) \leq E^U_r(R), \tag{18}$$

where,

$$E^U_r(R) = \max_{\rho\in[0,1]}\left\{E^U_0(\rho) - \rho R\right\},$$

and

The computation of $E^U_0(\rho)$, when $l = 1$, is done in [9]. Here, we do the computation for arbitrary $l$. The following lemma specifies an upper bound to $E^U_0(\rho)$:

**Lemma 3**

$$E^U_0(\rho) \leq rt\log\left(1 + \frac{\rho t\mathsf{SNR}^{-[2\nu-\min\{1,\nu\}]}}{(t+r)^2(1+\rho)}\right).$$

*Proof:* Since $\mathbf{H}$ is independent of $\mathbf{X}$,

$$p(Y,H|X) = p(H)p(Y|X,H).$$

Hence, $E^U_0(\rho)$ can be expressed as

$$E^U_0(\rho) = -\log\left(E_{\mathbf{H}}\left[\int\left[\int q(X)p(Y|X,\mathbf{H})^{\frac{1}{1+\rho}}dX\right]^{1+\rho}dY\right]\right).$$

The conditional probability $p(Y|X,H)$ is given by

$$p(Y|X,H) = \left(\frac{\mathsf{SNR}_b}{\pi t}\right)^{rl} \exp\left[-\frac{\mathsf{SNR}_b}{t}\mathsf{trace}\Big\{\Big(HX-Y\Big)^\dagger\Big(HX-Y\Big)\Big\}\right].$$

Defining $B$ as

$$\begin{aligned} B &\triangleq \frac{\mathsf{SNR}_b}{t(1+\rho)}H^\dagger H. \\ \Rightarrow B^{-1} &= \frac{t(1+\rho)}{\mathsf{SNR}_b}(H^\dagger)^{-1}(H)^{-1}. \end{aligned}$$

In the proof of this lemma, for any matrix $M$, we use $M^{-1}$ to denote its pseudoinverse.
Now,

$$\begin{aligned} &\int q(X)p(Y|X,H)^{\frac{1}{1+\rho}}dX \\ &\quad= \int \frac{1}{\pi^{lt}}\exp\Big[-\mathsf{trace}(X^\dagger X)\Big]\left(\frac{\mathsf{SNR}_b}{\pi t}\right)^{\frac{rl}{1+\rho}}\exp\Big[-\frac{\mathsf{SNR}_b}{t(1+\rho)}\mathsf{trace}\Big\{(HX-Y)^\dagger(HX-Y)\Big\}\Big]dX \\ &\quad= \frac{1}{\pi^{lt}}\left(\frac{\mathsf{SNR}_b}{\pi t}\right)^{\frac{rl}{1+\rho}}\int \exp\Big[-\mathsf{trace}\Big\{X^\dagger(I_t+B)X-\frac{\mathsf{SNR}_b}{t(1+\rho)}(X^\dagger H^\dagger Y+Y^\dagger HX-Y^\dagger Y)\Big\}\Big]dX \\ &\quad= \frac{1}{\pi^{lt}}\left(\frac{\mathsf{SNR}_b}{\pi t}\right)^{\frac{rl}{1+\rho}}\exp\Big[-\frac{\mathsf{SNR}_b}{t(1+\rho)}\mathsf{trace}\Big\{Y^\dagger(I_r-(I_r+B^{-1})^{-1})Y\Big\}\Big]\int\exp\Big[-\frac{\mathsf{SNR}_b}{t(1+\rho)} \\ &\qquad\qquad\cdot\mathsf{trace}\Big\{X^\dagger H^\dagger(B^{-1}+I_r)HX-X^\dagger H^\dagger Y-Y^\dagger HX+Y^\dagger(I_r+B^{-1})^{-1}Y\Big\}\Big]dX \\ &\quad= \left(\frac{\mathsf{SNR}_b}{\pi t}\right)^{\frac{rl}{1+\rho}}\exp\Big[-\frac{\mathsf{SNR}_b}{t(1+\rho)}\mathsf{trace}\Big\{Y^\dagger(I_r-(I_r+B^{-1})^{-1})Y\Big\}\Big]\det(I_t+B)^{-l} \end{aligned}$$

Therefore,

$$\begin{aligned} &\int\left[\int q(X)p(Y|X,H)^{\frac{1}{1+\rho}}dX\right]^{1+\rho}dY \\ &\quad= \left(\frac{\mathsf{SNR}_b}{\pi t}\right)^{rl}\det(I_t+B)^{-l(1+\rho)}\int\exp\Big[-\frac{\mathsf{SNR}_b}{t}\mathsf{trace}\Big\{Y^\dagger(I_r-(I_r+B^{-1})^{-1})Y\Big\}\Big]dY \\ &\quad= \det(I_t+B)^{-l(1+\rho)}\det\Big(I_r-(I_r+B^{-1})^{-1}\Big)^{-l} \\ &\quad= \det(I_t+B)^{-\rho l} \\ &\quad= \det\left(I_t+\frac{\mathsf{SNR}_b}{t(1+\rho)}H^\dagger H\right)^{-\rho l}. \end{aligned}$$

Hence,

$$\begin{aligned} &E_0^U(\rho) \\ &\quad= -\log E_{\mathbf{H}}\left[\det\left(I_t+\frac{\mathsf{SNR}_b}{t(1+\rho)}\mathbf{H}^\dagger\mathbf{H}\right)^{-\rho l}\right] \end{aligned} \tag{19}$$

$$
\begin{aligned}
&= -\log E_{\mathbf{H}}\left[\exp\left(-\rho l \log\det\left(I_t + \frac{\mathsf{SNR}_b}{t(1+\rho)}\mathbf{H}^\dagger\mathbf{H}\right)\right)\right] \\
&\leq -\log E_{\mathbf{H}}\left[\exp -\left(\frac{\rho l \mathsf{SNR}_b}{t(1+\rho)}\mathsf{trace}(\mathbf{H}^\dagger\mathbf{H})\right)\right] \qquad (20)\\
&= -\log \int_0^\infty \frac{x^{rt-1}}{(rt-1)!}\exp\left(-\left(1+\frac{\rho l \mathsf{SNR}_b}{t(1+\rho)}\right)x\right)dx \\
&= rt\log\left(1+\frac{\rho l \mathsf{SNR}_b}{t(1+\rho)}\right).
\end{aligned}
$$

To obtain (20), we use the following inequality:

$$
\begin{aligned}
&\log\det\left(I_t + \frac{\mathsf{SNR}_b}{t(1+\rho)}\mathbf{H}^\dagger\mathbf{H}\right) \\
&\quad = \sum_{i=1}^{\min(t,r)} \log\left(1+\frac{\mathsf{SNR}_b}{t(1+\rho)}\lambda_i(\mathbf{H}^\dagger\mathbf{H})\right) \\
&\quad \leq \frac{\mathsf{SNR}_b}{t(1+\rho)} \sum_{i=1}^{\min(t,r)} \lambda_i(\mathbf{H}^\dagger\mathbf{H}) \\
&\quad = \frac{\mathsf{SNR}_b}{t(1+\rho)}\mathsf{trace}(\mathbf{H}^\dagger\mathbf{H}).
\end{aligned}
$$

$\lambda_i(\mathbf{H}^\dagger\mathbf{H})$ is the $i^{th}$ eigenvalue of the random matrix $\mathbf{H}^\dagger\mathbf{H}$. Hence, $E_0^U(\rho)$ can be upper bounded as:

$$
\begin{aligned}
&E_0^U(\rho) \\
&\quad \leq rt\log\left(1+\frac{\rho l \mathsf{SNR}_b}{t(1+\rho)}\right) \\
&\quad = rt\log\left(1+\frac{\rho t \mathsf{SNR}^{-[2\nu-\min\{1,\nu\}]}}{(t+r)^2(1+\rho)}\right).
\end{aligned}
$$

This completes the proof of the lemma. □

Combining (18) with Lemma 3, we obtain an upper bound for $E_r(R)$:

$$
E_r(R) \leq \max_{\rho\in[0,1]}\left\{rt\log\left(1+\frac{\rho t \mathsf{SNR}^{-[2\nu-\min\{1,\nu\}]}}{(t+r)^2(1+\rho)}\right)-\rho R\right\}. \qquad (21)
$$

Since, $E_r(R)$ is positive over the rate range (11), any upper bound to it will also be positive over (11). In fact, since perfect knowledge of $\mathbf{H}$ at the receiver increases capacity, the upper bound is positive over a rate range larger than (11).

**Lower Bound to $E_r(R)$:**

We now use a training based scheme to obtain a lower bound on $E_r(R)$. Since this is one of the possible schemes that can be used for the non-coherent channel, the random coding error exponent for this scheme, $E_r^L(R)$, can be upper bounded as

$$
E_r^L(R) \leq E_r(R). \qquad (22)
$$

We rewrite the channel model within one coherence block as

$$\vec{\mathbf{y}}_i = \mathbf{H}\vec{\mathbf{x}}_i + \vec{\mathbf{w}}_i, \quad i = 1, \ldots, l. \tag{23}$$

The channel matrix, $\mathbf{H}$, is constant within the block. The total energy available in the block is:

$$\mathsf{E}_{\mathsf{total}} = l \cdot \mathsf{SNR}_b.$$

We use the first $t$ symbols of the block for training[3] using $\gamma \in (0, 1)$ fraction of the total energy. The remaining fraction is used for communicating data. Hence, the energy used for training is:

$$\mathsf{E}_{\mathsf{training}} = \gamma \mathsf{E}_{\mathsf{total}} = \gamma l \mathsf{SNR}_b.$$

The following training sequence is used:

$$[\vec{\mathbf{x}}_1 \ldots \vec{\mathbf{x}}_t] = \sqrt{\frac{\mathsf{E}_{\mathsf{training}}}{t}}.I_t.$$

This training scheme makes $\mathbf{y}_{i,j}$ a sufficient statistic for estimating $\mathbf{h}_{i,j}$. The receiver computes the minimum mean-squares error (MMSE) estimate of $\mathbf{H}$ from $[\vec{\mathbf{y}}_1 \ldots \vec{\mathbf{y}}_t]$. Using $\hat{\mathbf{h}}_{i,j}$ and $\tilde{\mathbf{h}}_{i,j}$ to denote the estimate and estimation error of $\mathbf{h}_{i,j}$, respectively, we have for $i \in \{1, \ldots, r\}, j \in \{1, \ldots, t\}$:

$$\hat{\mathbf{h}}_{i,j} \sim \mathcal{CN}\Big(0, \frac{\frac{\mathsf{E}_{\mathsf{training}}}{t}}{1 + \frac{\mathsf{E}_{\mathsf{training}}}{t}}\Big),$$
$$\tilde{\mathbf{h}}_{i,j} \sim \mathcal{CN}\Big(0, \frac{1}{1 + \frac{\mathsf{E}_{\mathsf{training}}}{t}}\Big),$$

and, $\hat{\mathbf{h}}_{i,j}, \tilde{\mathbf{h}}_{i,j}$ are independent due to the estimation being MMSE. Moreover, the sets $\{\hat{\mathbf{h}}_{i,j}\}$ and $\{\tilde{\mathbf{h}}_{i,j}\}$ have independent elements. Thus, representing the estimate and estimation error of the channel matrix as $\hat{\mathbf{H}}$ and $\tilde{\mathbf{H}}$, respectively, we have

$$\mathbf{H} = \hat{\mathbf{H}} + \tilde{\mathbf{H}},$$

where, $\hat{\mathbf{H}}$ and $\tilde{\mathbf{H}}$ are independent matrices, each with i.i.d Gaussian entries.
For the remaining $l - t$ symbols within the same block, $\mathsf{E}_{\mathsf{total}} - \mathsf{E}_{\mathsf{training}} = (1 - \gamma) l \mathsf{SNR}_b$ energy is used to send data using an i.i.d Gaussian code. The channel in this phase can be represented as

$$\vec{\mathbf{y}}_i = \hat{\mathbf{H}}\vec{\mathbf{x}}_i + \underbrace{\tilde{\mathbf{H}}\vec{\mathbf{x}}_i + \vec{\mathbf{w}}_i}_{\vec{\mathbf{v}}_i}, \qquad i = t + 1, \ldots, l. \tag{24}$$

$\{\vec{\mathbf{x}}_i\}$ are i.i.d complex Gaussian vectors:

$$\vec{\mathbf{x}}_i \sim \mathcal{CN}\left(0, \frac{(1 - \gamma) l \mathsf{SNR}_b}{(l - t) t} \cdot I_t\right).$$

[3] We assume $l > t$.

$\tilde{\mathbf{H}}\vec{\mathbf{x}}_i$ is the noise due to the estimation error from the training phase coupled with the input signal. Combining the additive white noise with the noise due to estimation error, we have

$$\vec{\mathbf{v}}_i \triangleq \tilde{\mathbf{H}}\vec{\mathbf{x}}_i + \vec{\mathbf{w}}_i.$$

Note that $\vec{\mathbf{v}}_i$ is uncorrelated, but not independent of $\hat{\mathbf{H}}\vec{\mathbf{x}}_i$. Its covariance matrix is

$$\begin{aligned}
&E[\vec{\mathbf{v}}_i\vec{\mathbf{v}}_i^\dagger] \\
&\quad = E[\tilde{\mathbf{H}}\vec{\mathbf{x}}_i\vec{\mathbf{x}}_i^\dagger\tilde{\mathbf{H}}^\dagger] + I_r \\
&\quad = \left[\frac{(1-\gamma)l\mathsf{SNR}_b}{t(l-t)}\cdot\frac{t}{1+\frac{\mathsf{E}_{\mathsf{training}}}{t}} + 1\right]\cdot I_r \\
&\quad = \left[\frac{t(1-\gamma)l\mathsf{SNR}_b}{(l-t)(t+\gamma l\mathsf{SNR}_b)} + 1\right]\cdot I_r.
\end{aligned}$$

The channel in (24) can be normalized to:

$$\vec{\mathbf{y}}_i = \mathbf{H}^{'}\vec{\mathbf{x}}_i + \vec{\mathbf{v}}_i^{'}, \quad i = t+1,\ldots,l, \tag{25}$$

where, $\mathbf{H}^{'}$ has i.i.d $\mathcal{C}N(0,1)$ entries and is perfectly known at the receiver (this is the MMSE estimate) , $\vec{\mathbf{v}}_i^{'}$ is a zero-mean noise vector having the covariance matrix

$$E[\vec{\mathbf{v}}_i^{'}\vec{\mathbf{v}}_i^{'\dagger}] = I_r,$$

and, $\{\vec{\mathbf{x}}_i\}$ are i.i.d complex Gaussian vectors:

$$\vec{\mathbf{x}}_i \sim \mathcal{CN}(0, \frac{f(\gamma,\mathsf{SNR})}{t}\cdot I_t),$$

where

$$f(\gamma,\mathsf{SNR}) = \frac{\frac{\gamma l\mathsf{SNR}_b}{t+\gamma l\mathsf{SNR}_b}\cdot\frac{(1-\gamma)l\mathsf{SNR}_b}{(l-t)}}{\frac{t(1-\gamma)l\mathsf{SNR}_b}{(l-t)(t+\gamma l\mathsf{SNR}_b)}+1}.$$

Now,

$$\begin{aligned}
&f(\gamma,\mathsf{SNR}) \\
&\quad = l\mathsf{SNR}_b^2\cdot\frac{\gamma(1-\gamma)l}{t(1-\gamma)l\mathsf{SNR}_b + (l-t)(t+\gamma l\mathsf{SNR}_b)} \\
&\quad \geq l\mathsf{SNR}_b^2\cdot\frac{\gamma(1-\gamma)}{\gamma l\mathsf{SNR}_b + t(1+\mathsf{SNR}_b)} \\
&\quad \geq \frac{l\mathsf{SNR}_b^2}{t(1+\mathsf{SNR}_b)}\cdot\frac{\gamma(1-\gamma)}{1+\gamma\frac{l\mathsf{SNR}_b}{t(1+\mathsf{SNR}_b)}}.
\end{aligned} \tag{26}$$

Define

$$f^*(\mathsf{SNR}) \triangleq \max_{\gamma\in(0,1)} f(\gamma,\mathsf{SNR}).$$

Using (26), we get a lower bound to $f^*(\mathsf{SNR})$:

$$
\begin{aligned}
& f^*(\mathsf{SNR}) \\
& \quad \geq \frac{l\mathsf{SNR}_b^2}{t(1+\mathsf{SNR}_b)} \cdot \max_{\gamma\in(0,1)} \left\{ \frac{\gamma(1-\gamma)}{1+\gamma\frac{l\mathsf{SNR}_b}{t(1+\mathsf{SNR}_b)}} \right\} \\
& \quad = \mathsf{SNR}_b \left[ 1 - \frac{2\sqrt{1+\frac{l\mathsf{SNR}_b}{t(1+\mathsf{SNR}_b)}}}{\frac{l\mathsf{SNR}_b}{t(1+\mathsf{SNR}_b)}} + \frac{2}{\frac{l\mathsf{SNR}_b}{t(1+\mathsf{SNR}_b)}} \right] \\
& \quad = \mathsf{SNR}^{\min\{1,\nu\}} - 2\frac{(r+t)}{\sqrt{t}}\mathsf{SNR}^{\nu+\frac{\min\{1,\nu\}}{2}} + o\Big(\mathsf{SNR}^{\nu+\frac{\min\{1,\nu\}}{2}}\Big) \triangleq f^*_{LB}(\mathsf{SNR}).
\end{aligned}
$$

Hence,

$$
f^*(\mathsf{SNR}) \geq f^*_{LB}(\mathsf{SNR}) = \mathsf{SNR}^{\min\{1,\nu\}} - 2\frac{(t+r)}{\sqrt{t}}\mathsf{SNR}^{\nu+\frac{\min\{1,\nu\}}{2}} + o\Big(\mathsf{SNR}^{\nu+\frac{\min\{1,\nu\}}{2}}\Big). \tag{27}
$$

Note that

$$
f^*_{LB}(\mathsf{SNR}) = \mathsf{SNR}^{\min\{1,\nu\}} + o(\mathsf{SNR}^{\min\{1,\nu\}}).
$$

The random coding error exponent for this scheme is

$$
E_r^L(R) = \max_{\rho\in[0,1]} \left\{ E_0^L(\rho) - \rho R \right\}, \tag{28}
$$

where,

Since the training and data communication phases use independent input signals, $\mathbf{H}'$ is independent of $\mathbf{X}$. Thus

$$
p(Y, H'|X) = p(H')p(Y|X, H'),
$$

and

The following lemma specifies a lower bound to $E_0^L(\rho)$:

**Lemma 4**

$$
E_0^L(\rho) \geq rt\log\left(1 + \frac{\rho t \mathsf{SNR}^{-[2\nu-\min\{1,\nu\}]}}{(t+r)^2(1+\rho)}\right) - o(1).
$$

*Proof:* References [11, 19] show that capacity of the channel in (25) is minimized if $\{\vec{\mathbf{v}}_i\}$ are i.i.d Gaussian:

$$
\vec{\mathbf{v}}_i \sim \mathcal{CN}(0, I_r) \quad i = t+1, \ldots, l.
$$

We conjecture that this noise distribution also minimizes error exponent. With this assumption, the error exponent for this channel with i.i.d Gaussian noise is similar to that of the coherent channel ($E_0^U(\rho)$ in (19), with $\mathsf{SNR}_b$ replaced by $f(\gamma, \mathsf{SNR})$ and $l$ replaced by $l-t$). Hence, we obtain the following lower bound:

$$
\begin{aligned}
& E_0^L(\rho) \\
& \geq \max_{\gamma\in(0,1)} \left\{ -\log E_{\mathbf{H}'} \left[ \det\left( I_t + \frac{f(\gamma, \mathsf{SNR})}{t(1+\rho)} \mathbf{H}'^{\dagger}\mathbf{H}' \right)^{-\rho(l-t)} \right] \right\} \\
& = -\log E_{\mathbf{H}'} \left[ \det\left( I_t + \frac{\max_{\gamma\in(0,1)} f(\gamma, \mathsf{SNR})}{t(1+\rho)} \mathbf{H}'^{\dagger}\mathbf{H}' \right)^{-\rho(l-t)} \right] \\
& = -\log E_{\mathbf{H}'} \left[ \det\left( I_t + \frac{f^*(\mathsf{SNR})}{t(1+\rho)} \mathbf{H}'^{\dagger}\mathbf{H}' \right)^{-\rho(l-t)} \right] \\
& \geq -\log E_{\mathbf{H}'} \left[ \det\left( I_t + \frac{f^*_{LB}(\mathsf{SNR})}{t(1+\rho)} \mathbf{H}'^{\dagger}\mathbf{H}' \right)^{-\rho(l-t)} \right] \\
& \geq -\log E_{\mathbf{H}'} \left[ \left( 1 + \frac{f^*_{LB}(\mathsf{SNR})}{t(1+\rho)} \mathsf{trace}(\mathbf{H}'^{\dagger}\mathbf{H}') \right)^{-\rho(l-t)} \right] \qquad (29) \\
& = -\log \int_0^\infty \frac{x^{rt-1}\exp(-x)}{(rt-1)!} \left( 1 + \frac{f^*_{LB}(\mathsf{SNR})}{t(1+\rho)} x \right)^{-\rho(l-t)} dx \\
& = -\log \int_0^\infty \frac{x^{rt-1}}{(rt-1)!} \exp\left[ -x - \rho(l-t)\log\left( 1 + \frac{f^*_{LB}(\mathsf{SNR})}{t(1+\rho)} x \right) \right] dx \\
& = -\log C, \qquad (30)
\end{aligned}
$$

where,

$$
C \triangleq \int_0^\infty \frac{x^{rt-1}}{(rt-1)!} \exp\left[ -x - \rho(l-t)\log\left( 1 + \frac{f^*_{LB}(\mathsf{SNR})}{t(1+\rho)} x \right) \right] dx.
$$

Equation (29) holds due the following inequality:

$$
\begin{aligned}
& \det\left( I_t + \frac{f^*_{LB}(\mathsf{SNR})}{t(1+\rho)} \mathbf{H}'^{\dagger}\mathbf{H}' \right) \\
& \quad = \prod_{i=1}^{\min(t,r)} \left( 1 + \frac{f^*_{LB}(\mathsf{SNR})}{t(1+\rho)} \lambda_i(\mathbf{H}'^{\dagger}\mathbf{H}') \right) \\
& \quad \geq 1 + \frac{f^*_{LB}(\mathsf{SNR})}{t(1+\rho)} \sum_{i=1}^{\min(t,r)} \lambda_i(\mathbf{H}'^{\dagger}\mathbf{H}') \\
& \quad = 1 + \frac{f^*_{LB}(\mathsf{SNR})}{t(1+\rho)} \mathsf{trace}(\mathbf{H}'^{\dagger}\mathbf{H}'). \qquad (31)
\end{aligned}
$$

$\lambda_i(\mathbf{H}'^{\dagger}\mathbf{H}')$ is the $i^{th}$ eigenvalue of the random matrix $\mathbf{H}'^{\dagger}\mathbf{H}'$.
We now compute an upper bound to $C$. Splitting the range of integration, we have

$$
C = C_1 + C_2, \qquad (32)
$$

where,

$$C_1 = \int_0^{2t(1+\rho)} \frac{x^{rt-1}}{(rt-1)!} \exp\left[-x - \rho(l-t)\log\left(1+\frac{f_{LB}^*(\mathsf{SNR})}{t(1+\rho)}x\right)\right] dx,$$
$$C_2 = \int_{2t(1+\rho)}^{\infty} \frac{x^{rt-1}}{(rt-1)!} \exp\left[-x - \rho(l-t)\log\left(1+\frac{f_{LB}^*(\mathsf{SNR})}{t(1+\rho)}x\right)\right] dx.$$

$C_1$ can be upper bounded as:

$$\begin{aligned}
&C_1 \\
&\leq \int_0^{2t(1+\rho)} \frac{x^{rt-1}}{(rt-1)!} \exp\left[-x - \frac{\rho(l-t)f_{LB}^*(\mathsf{SNR})x}{t(1+\rho)}\left(1-\frac{f_{LB}^*(\mathsf{SNR})}{2t(1+\rho)}x\right)\right] dx \\
&\leq \int_0^{2t(1+\rho)} \frac{x^{rt-1}}{(rt-1)!} \exp\left[-x - \frac{\rho(l-t)f_{LB}^*(\mathsf{SNR})(1-f_{LB}^*(\mathsf{SNR}))}{t(1+\rho)}x\right] dx \\
&\leq \int_0^{\infty} \frac{x^{rt-1}}{(rt-1)!} \exp\left[-x - \frac{\rho(l-t)f_{LB}^*(\mathsf{SNR})(1-f_{LB}^*(\mathsf{SNR}))}{t(1+\rho)}x\right] dx \\
&= \left[1+\frac{\rho(l-t)f_{LB}^*(\mathsf{SNR})(1-f_{LB}^*(\mathsf{SNR}))}{t(1+\rho)}\right]^{-rt} \\
&= \left[1+\frac{\rho}{t(1+\rho)}\left[\left(l-t\right)\left(\mathsf{SNR}^{\min\{1,\nu\}}+o(\mathsf{SNR}^{\min\{1,\nu\}})\right)\left(1-\mathsf{SNR}^{\min\{1,\nu\}}+o(\mathsf{SNR}^{\min\{1,\nu\}})\right)\right]\right]^{-rt} \\
&= \left[1+\frac{\rho t\mathsf{SNR}^{-[2\nu-\min\{1,\nu\}]}}{(t+r)^2(1+\rho)}\left[(1-\frac{(t+r)^2}{t}\mathsf{SNR}^{2\nu})(1+o(1))(1-\mathsf{SNR}^{\min\{1,\nu\}}+o(\mathsf{SNR}^{\min\{1,\nu\}}))\right]\right]^{-rt} \\
&= \left[1+\frac{\rho t\mathsf{SNR}^{-[2\nu-\min\{1,\nu\}]}}{(t+r)^2(1+\rho)}\left(1+o(1)\right)\right]^{-rt} \\
&= \left[1+\frac{\rho t\mathsf{SNR}^{-[2\nu-\min\{1,\nu\}]}}{(t+r)^2(1+\rho)}\right]^{-rt}\left[1+o(1)\right]. \qquad (33)
\end{aligned}$$

$C_2$ can be upper bounded as:

$$\begin{aligned}
&C_2 \\
&\leq \int_{2t(1+\rho)}^{\infty} \frac{x^{rt-1}}{(rt-1)!} \exp\left[-x - \rho(l-t)\log\left(1+2f_{LB}^*(\mathsf{SNR})\right)\right] dx \\
&\leq \int_0^{\infty} \frac{x^{rt-1}}{(rt-1)!} \exp\left[-x - \rho(l-t)\log\left(1+2f_{LB}^*(\mathsf{SNR})\right)\right] dx \\
&= \exp\left[-\rho(l-t)\log\left(1+2f_{LB}^*(\mathsf{SNR})\right)\right]. \qquad (34)
\end{aligned}$$

Combining (32, 33, 34), we get the upper bound for $C$ as:

$$\begin{aligned}
&C \qquad (35) \\
&\leq \left[1+\frac{\rho t\mathsf{SNR}^{-[2\nu-\min\{1,\nu\}]}}{(t+r)^2(1+\rho)}\right]^{-rt}\left[1+o(1)\right] + \exp\left[-\rho(l-t)\log\left(1+2f_{LB}^*(\mathsf{SNR})\right)\right]
\end{aligned}$$

$$= \left[1 + \frac{\rho t \mathsf{SNR}^{-[2\nu - \min\{1,\nu\}]}}{(t+r)^2(1+\rho)}\right]^{-rt} \Big[1 + o(1)\Big]\Big[1 + \Big[1 + o(1)\Big]\Big[1 + \frac{\rho t \mathsf{SNR}^{-[2\nu - \min\{1,\nu\}]}}{(t+r)^2(1+\rho)}\Big]^{rt}$$

$$\cdot \exp\Big[-\rho\Big(\frac{t^2}{(t+r)^2}\mathsf{SNR}^{-2\nu} - t\Big)\log(1 + 2(\mathsf{SNR}^{\min\{1,\nu\}} + o(\mathsf{SNR}^{\min\{1,\nu\}})))\Big]\Big]$$

$$= \left[1 + \frac{\rho t \mathsf{SNR}^{-[2\nu - \min\{1,\nu\}]}}{(t+r)^2(1+\rho)}\right]^{-rt} \Big[1 + o(1)\Big]\Big[1 + \Big[1 + o(1)\Big] \cdot o(1)\Big]$$

$$= \left[1 + \frac{\rho t \mathsf{SNR}^{-[2\nu - \min\{1,\nu\}]}}{(t+r)^2(1+\rho)}\right]^{-rt} \Big[1 + o(1)\Big]. \tag{36}$$

From (30, 36), we get a lower bound to $E_0^L(\rho)$ as:

$$E_0^L(\rho) \geq rt \log\Big(1 + \frac{\rho t \mathsf{SNR}^{-[2\nu - \min\{1,\nu\}]}}{(t+r)^2(1+\rho)}\Big) - o(1). \tag{37}$$

This completes the proof of the lemma. □

Combining (22, 28) with Lemma 4, we obtain a lower bound for $E_r(R)$:

$$E_r(R) \geq \max_{\rho \in [0,1]} \left\{ rt \log\left(1 + \frac{\rho t \mathsf{SNR}^{-[2\nu - \min\{1,\nu\}]}}{(t+r)^2(1+\rho)}\right) - \rho R \right\} - o(1). \tag{38}$$

Since the training based scheme has a lower capacity than the non-coherent capacity [11, 19], the range of rates for which the error exponent for the training based scheme is positive, is reduced from (11). We compute a lower bound to the capacity for this scheme.

Letting $\{\vec{\mathbf{v}}_i'\}$ be i.i.d white Gaussian vectors, i.e., $\vec{\mathbf{v}}_i' \sim \mathcal{CN}(0, I_r)$ in (25), we can lower bound the capacity per block used for transmission, $C_{\mathsf{T}}^{\mathsf{block}}(\mathsf{SNR})$, as

$$\begin{aligned}
&C_{\mathsf{T}}^{\mathsf{block}}(\mathsf{SNR}) \\
&\quad \geq (l-t).\max_{\gamma \in (0,1)} \left\{ \log\det\left[I_t + \frac{f(\gamma, \mathsf{SNR})}{t}\mathbf{H}'^{\dagger}\mathbf{H}'\right]\right\} \\
&\quad = (l-t).\log\det\left[I_t + \frac{\max_{\gamma\in(0,1)} f(\gamma, \mathsf{SNR})}{t}\mathbf{H}'^{\dagger}\mathbf{H}'\right] \\
&\quad = (l-t).\log\det\left[I_t + \frac{f^*(\mathsf{SNR})}{t}\mathbf{H}'^{\dagger}\mathbf{H}'\right] \\
&\quad \geq (l-t).\log\det\left[I_t + \frac{f^*_{LB}(\mathsf{SNR})}{t}\mathbf{H}'^{\dagger}\mathbf{H}'\right] \\
&\quad \geq (l-t).\left[r f^*_{LB}(\mathsf{SNR}) - \frac{r(r+t)}{2t} f^{*2}_{LB}(\mathsf{SNR})\right] \\
&\quad \geq \frac{t^2}{(t+r)^2}\mathsf{SNR}^{-2\nu}\Big[r\mathsf{SNR}^{\min\{1,\nu\}} - 2\frac{r(r+t)}{\sqrt{t}}\mathsf{SNR}^{\nu + \frac{\min\{1,\nu\}}{2}} \\
&\qquad\qquad - \frac{r(r+t)}{2t}\mathsf{SNR}^{2\min\{1,\nu\}} + o\Big(\mathsf{SNR}^{\min\{\nu + \frac{\min\{1,\nu\}}{2}, 2\min\{1,\nu\}\}}\Big)\Big] \triangleq C_{\mathsf{T,lb}}^{\mathsf{block}}(\mathsf{SNR}).
\end{aligned} \tag{39}$$

Hence, the lower bound to $E_r(R)$ in (38) is positive in the range

$$0 \leq R \leq C^{\mathsf{block}}_{\mathsf{T,lb}}(\mathsf{SNR}).$$

**Combining the upper and lower bounds**:
Combining (21,38), we have

$$\begin{aligned} E_r(R) &= \max_{\rho\in[0,1]} \left\{ rt\log\left(1+\frac{\rho t\mathsf{SNR}^{-[2\nu-\min\{1,\nu\}]}}{(t+r)^2(1+\rho)}\right) - \rho R\right\} - o(1), \\ 0 &\leq R \leq C^{\mathsf{block}}_{\mathsf{T,lb}}(\mathsf{SNR}). \end{aligned} \tag{40}$$

Let

$$\rho^* = \arg\max_{\rho\in[0,1]} \left\{ rt\log\left(1+\frac{\rho t\mathsf{SNR}^{-[2\nu-\min\{1,\nu\}]}}{(t+r)^2(1+\rho)}\right) - \rho R\right\}.$$

We compute $\rho^*$ as:

$$\rho^* = \begin{cases} 1 & 0 \leq R \leq R_{\mathsf{critical}} \\ \frac{1}{2}\left[\sqrt{1+4\left(\frac{rt}{R} - \frac{(t+r)^2\mathsf{SNR}^{2\nu-\min\{1,\nu\}}}{t}\right)} - 1\right] & R_{\mathsf{critical}} \leq R \leq C^{\mathsf{block}}_{\mathsf{T,lb}}(\mathsf{SNR}) \end{cases}$$

where, $R_{\mathsf{critical}}$, the critical rate, is

$$R_{\mathsf{critical}} = rt/2 + o(1).$$

Substituting $\rho^*$ in (40), we have for $0 \leq R \leq R_{\mathsf{critical}}$,

$$E_r(R) = rt\log\left(1+\frac{t\mathsf{SNR}^{-[2\nu-\min\{1,\nu\}]}}{2(t+r)^2}\right) - R - o(1), \tag{41}$$

and, for $R_{\mathsf{critical}} \leq R \leq C^{\mathsf{block}}_{\mathsf{T,lb}}(\mathsf{SNR})$,

$$\begin{aligned} &E_r(R) \\ &= rt\log\left(1+\left[\frac{\sqrt{1+4\left(\frac{rt}{R}-\frac{(t+r)^2\mathsf{SNR}^{2\nu-\min\{1,\nu\}}}{t}\right)}-1}{\sqrt{1+4\left(\frac{rt}{R}-\frac{(t+r)^2\mathsf{SNR}^{2\nu-\min\{1,\nu\}}}{t}\right)}+1}\right]\left[\frac{t\mathsf{SNR}^{-[2\nu-\min\{1,\nu\}]}}{(t+r)^2}\right]\right) \\ &\qquad - \frac{R}{2}\left[\sqrt{1+4\left(\frac{rt}{R}-\frac{(t+r)^2\mathsf{SNR}^{2\nu-\min\{1,\nu\}}}{t}\right)}-1\right] - o(1). \end{aligned} \tag{42}$$

For $R \in \left(C^{\mathsf{block}}_{\mathsf{T,lb}}(\mathsf{SNR}), C^{\mathsf{block}}(\mathsf{SNR})\right]$, the lower bound (38) to $E_r(R)$, is 0. However, the upper bound (21) is $o(1)$ in this range. Hence, we can say that

$$E_r(R) = o(1) \text{ for } C^{\mathsf{block}}_{\mathsf{T,lb}}(\mathsf{SNR}) \leq R \leq C^{\mathsf{block}}(\mathsf{SNR}). \tag{43}$$

Equations(41-43) together characterize the random coding error exponent for the non-coherent channel. This completes the proof of Theorem 4.

# 5 Conclusions

In this paper, we have computed the capacity and error probability for the non-coherent wideband MIMO channel. The effect on capacity and reliability of coherence length and number of transmit and receive antennas have been examined. The analysis has shown that though the number of transmit antennas does not effect the linear capacity term, it does effect the sublinear capacity term, i.e., the approach of capacity to the wideband limit with increasing bandwidth. We have also established conditions on the channel coherence length and number of antennas, for the non-coherent capacity to be the same as the coherent capacity in the wideband regime. The error probability is shown to decay inversely with coherence length and exponentially with product of the number of transmit and receive antennas. This highlights the importance of multiple transmit antennas, besides multiple receive antennas, in the low SNR regime. An interesting observation has been that outage probability dominates the error probability even at low SNR.

# Appendix 1

**Proof of Lemma 1:**
*Proof of (2):* For any $\alpha \in (0,1]$ and $\gamma \in (0,\alpha)$, let there exist an input distribution on $\mathbf{X}$ that satisfies the following:

$$\frac{1}{l} I(\mathbf{X};\mathbf{Y}) \geq r\mathsf{SNR} - \frac{r(r+t)}{2t}\mathsf{SNR}^{1+\alpha} + O(\mathsf{SNR}^{1+\alpha+\gamma}). \tag{44}$$

Let $\mathbf{Y^G}$ be a matrix, with i.i.d complex Gaussian entries, that satisfies the same power constraint as the received matrix, $\mathbf{Y}$, i.e. $E[\mathsf{trace}(\mathbf{Y^G}\mathbf{Y^{G\dagger}})] = E[\mathsf{trace}(\mathbf{Y}\mathbf{Y}^\dagger)]$. This makes $h(\mathbf{Y}) \leq h(\mathbf{Y^G})$ and the entries of $\mathbf{Y^G}$ i.i.d $\mathcal{CN}(0, 1+\mathsf{SNR})$. Moreover, conditioned on $\mathbf{X}$, the row vectors of $\mathbf{Y}$ are i.i.d $\mathcal{CN}(0, \mathbf{X}\mathbf{X}^\dagger + I_l)$. We can thus upper bound the mutual information as

$$\begin{aligned} & I(\mathbf{X};\mathbf{Y}) \\ & \quad = h(\mathbf{Y}) - h(\mathbf{Y}|\mathbf{X}) \\ & \quad \leq h(\mathbf{Y^G}) - h(\mathbf{Y}|\mathbf{X}) \\ & \quad = rl\log(1+\mathsf{SNR}) - rtE[\log(1+\|\vec{\mathbf{x}}_i^T\|^2)] \\ & \quad \leq rl\mathsf{SNR} - rtE[\log(1+\|\vec{\mathbf{x}}_i^T\|^2)] \end{aligned} \tag{45}$$

Combining (44) and (45) and noting that the norms of the input vectors $\|\vec{\mathbf{x}}_i^T\|$ are identically distributed, we see that if the input distribution satisfies (44), then it necessarily satisfies the first condition (2).
*Proof of (3):* Observing the structure of the optimal input [8] for the non-coherent MIMO channel, we can upper bound the mutual information as

$$I(\mathbf{X};\mathbf{Y}) \leq I(\mathbf{A};\mathbf{Y}|\mathbf{\Phi}) + I(\mathbf{\Phi};\mathbf{Y}|\mathbf{A}), \tag{46}$$

where, $I(\mathbf{A};\mathbf{Y}|\mathbf{\Phi})$ is the information conveyed by the norm of the transmitted signal vectors given that the receiver has side information about their directions, and $I(\mathbf{\Phi};\mathbf{Y}|\mathbf{A})$ is the information conveyed by the direction of these vectors when the receiver has side information about their norm. We establish upper bounds on these two terms.

**Upper bound for $I(\mathbf{A};\mathbf{Y}|\mathbf{\Phi})$:**
When the receiver has side information about $\mathbf{\Phi}$, it can filter out noise orthogonal to the subspace spanned by the row vectors of $\mathbf{\Phi}$ to obtain an equivalent channel

$$\begin{aligned}
&\mathbf{Y\Phi}^\dagger \\
&\quad = \mathbf{HX\Phi}^\dagger + \mathbf{W\Phi}^\dagger \\
&\quad = \mathbf{HA} + \mathbf{W}^{'},
\end{aligned}$$

where $\mathbf{W}^{'}$ has the same distribution as $\mathbf{W}$ and there is no loss in information since $\mathbf{Y\Phi}^\dagger$ is a sufficient statistic for estimating $\mathbf{X}$ from $\mathbf{Y}$. Therefore

$$\begin{aligned}
&I(\mathbf{A};\mathbf{Y}|\mathbf{\Phi}) \\
&\quad = I(\mathbf{A};\mathbf{Y\Phi}^\dagger|\mathbf{\Phi}) \\
&\quad = I(\mathbf{A};\mathbf{HA}+\mathbf{W}^{'}|\mathbf{\Phi}) \\
&\quad \leq \sum_{i=1}^{t} I(\|\vec{\mathbf{x}}_i^T\|;\vec{\mathbf{h}}_i\|\vec{\mathbf{x}}_i^T\| + \vec{\mathbf{w}}_i^{'T}|\vec{\phi}_i^T) \\
&\quad \leq \sum_{i=1}^{t}\sum_{j=1}^{r} I(\|\vec{\mathbf{x}}_i^T\|;\mathbf{h}_{ij}\|\vec{\mathbf{x}}_i^T\| + \mathbf{w}_{ij}^{'}|\vec{\phi}_i^T),
\end{aligned} \tag{47}$$

where the last two inequalities follow from the chain rule of mutual information and the fact that conditioning reduces entropy. In order to get an upper bound on $I(\|\vec{\mathbf{x}}_i^T\|;\mathbf{h}_{ij}\|\vec{\mathbf{x}}_i^T\| + \mathbf{w}_{ij}^{'}|\vec{\phi}_i^T)$, we need to maximize this mutual information with the average power constraint $\frac{l}{t}\mathsf{SNR}$ and the constraint specified by (2). If we relax the latter constraint (2), then the mutual information is that of a single-input, single-output i.i.d Rayleigh fading channel with average power constraint $\frac{l}{t}\mathsf{SNR}$. From [13], we know that this mutual information is maximized by an on-off distribution of the form

$$\|\vec{\mathbf{x}}_i^T\|^2 = \begin{cases} \frac{l}{\zeta}\frac{\mathsf{SNR}}{t} & w.p. \quad \zeta \\ 0 & w.p. \quad 1-\zeta \end{cases}$$

for $\forall i \in \{1,\ldots,t\}$ and some $\zeta > 0$. This signalling scheme becomes increasingly "flashy" as the SNR gets low, i.e., $\zeta \to 0$ as $\mathsf{SNR} \to 0$. Hence, (47) becomes

$$\begin{aligned}
&I(\mathbf{A};\mathbf{Y}|\mathbf{\Phi}) \\
&\quad \leq \sum_{i=1}^{t}\sum_{j=1}^{r} I(\|\vec{\mathbf{x}}_i^T\|;\mathbf{h}_{ij}\|\vec{\mathbf{x}}_i^T\| + \mathbf{w}_{ij}^{'}|\vec{\phi}_i^T) \\
&\quad \leq \sum_{i=1}^{t}\sum_{j=1}^{r} H(\|\vec{\mathbf{x}}_i^T\|) \\
&\quad \approx rt\zeta\log(\frac{1}{\zeta}),
\end{aligned}$$

where the approximation is valid since we are in the low signal to noise ratio regime and $\zeta \to 0$ as $\mathsf{SNR} \to 0$. Therefore, we have

$$\frac{1}{l}I(\mathbf{A};\mathbf{Y}|\mathbf{\Phi}) \leq \frac{rt\zeta}{l}\log(\frac{1}{\zeta}). \tag{48}$$

However, this on-off distribution minimizes (2) also and hence the extra constraint does not change the optimal input. Therefore, it suffices to consider on-off signals. Hence, (2) becomes

$$\begin{aligned}
&\frac{(r+t)}{2t}\mathsf{SNR}^{1+\alpha} + O(\mathsf{SNR}^{1+\alpha+\gamma}) \\
&\quad \geq \frac{t\zeta}{l}\log(1+\frac{l}{\zeta t}\mathsf{SNR}) \\
&\quad \geq \frac{t\zeta}{l}\Big[\log(\frac{1}{\zeta}) - \log(\frac{t}{l\mathsf{SNR}})\Big] \\
&\quad \approx \frac{t\zeta}{l}\log(\frac{1}{\zeta}),
\end{aligned} \tag{49}$$

where the approximation is valid since $\frac{l}{\zeta}\frac{\mathsf{SNR}}{t} \gg 1$ as $\mathsf{SNR} \to 0$, i.e. the peak amplitude becomes very large as the signal to noise ratio tends to 0. Combining (48) and (49), we have

$$\frac{1}{l}I(\mathbf{A};\mathbf{Y}|\mathbf{\Phi}) \leq \frac{r(r+t)}{2t}\mathsf{SNR}^{1+\alpha} + O(\mathsf{SNR}^{1+\alpha+\gamma}). \tag{50}$$

**Upper bound for $I(\mathbf{\Phi};\mathbf{Y}|\mathbf{A})$:**

We can upper bound $I(\mathbf{\Phi};\mathbf{Y}|\mathbf{A})$ in terms of the mutual information of a single-input, single-output channel, i.e.

$$I(\mathbf{\Phi};\mathbf{Y}|\mathbf{A}) \leq \sum_{i=1}^{t}\sum_{j=1}^{r} I(\vec{\phi}_i^T;\vec{\mathbf{y}}_j^T|\mathbf{A},\vec{\phi}_1^T,\ldots,\vec{\phi}_{i-1}^T,\vec{\phi}_{i+1}^T,\ldots,\vec{\phi}_t^T,\vec{\mathbf{y}}_1^T,\ldots,\vec{\mathbf{y}}_{j-1}^T,\vec{\mathbf{y}}_{j+1}^T,\ldots,\vec{\mathbf{y}}_r^T).$$

The term inside the double summation represents the mutual information of the channel between the $i^{th}$ transmit antenna and $j^{th}$ receive antenna when no other antenna is present and the norm of $\vec{\mathbf{x}}_{\mathbf{i}}^{\mathbf{T}}$ is known at the receiver. Since the input vectors are identically distributed and the channel matrix has i.i.d. entries, the mutual information between any pair of transmit and receive antennas given that the other antennas are absent will be the same. Hence for all $i \in \{1,\ldots,t\}$ and $j \in \{1,\ldots,r\}$,

$$I(\mathbf{\Phi};\mathbf{Y}|\mathbf{A}) \leq rtI(\vec{\phi}_i^T;\vec{\mathbf{y}}_j^T|\mathbf{A},\vec{\phi}_1^T,\ldots,\vec{\phi}_{i-1}^T,\vec{\phi}_{i+1}^T,\ldots,\vec{\phi}_t^T,\vec{\mathbf{y}}_1^T,\ldots,\vec{\mathbf{y}}_{j-1}^T,\vec{\mathbf{y}}_{j+1}^T,\ldots,\vec{\mathbf{y}}_r^T).$$

We may thus consider the single-input, single-output channel between the $i^{th}$ transmit antenna and $j^{th}$ receive antenna:

$$\vec{\mathbf{y}}_j^T = \mathbf{h}_{ij}\|\vec{\mathbf{x}}_i^T\| + \vec{\mathbf{w}}_j^T.$$

Hence,

$$\begin{aligned}
&I(\mathbf{\Phi};\mathbf{Y}|\mathbf{A}) \\
&\quad \leq rtI(\vec{\mathbf{x}}_i^T;\vec{\mathbf{y}}_j^T|\|\vec{\mathbf{x}}_i^T\|) \\
&\quad = rtEI(\vec{\mathbf{x}}_i^T;\vec{\mathbf{y}}_j^T|\|\vec{x}_i^T\|).
\end{aligned} \tag{51}$$

Since $I(\vec{\mathbf{x}}_i^T; \mathbf{y}_j^T | \|\vec{x}_i^T\|)$ is the mutual information of a single-input, single-output channel over $l$ channel uses, it has a power constraint of $\frac{\|\vec{x}_i^T\|}{l}$. This mutual information can be upper bounded by the capacity of AWGN channel with the same power constraint, i.e.

$$I(\vec{\mathbf{x}}_i^T; \vec{\mathbf{y}}_j^T | \|\vec{x}_i^T\|) \leq l \log(1 + \frac{\|\vec{x}_i^T\|}{l}). \tag{52}$$

Combining (51) with (52), we obtain

$$I(\mathbf{\Phi}; \mathbf{Y}|\mathbf{A}) \leq rtlE \log(1 + \frac{\|\vec{\mathbf{x}}_i^T\|}{l}) \tag{53}$$

From (58), (50) and (53), we obtain our upper bound to $I(\mathbf{X}; \mathbf{Y})$ as

$$\frac{1}{l} I(\mathbf{X}; \mathbf{Y}) \leq rtE \log(1 + \frac{\|\vec{\mathbf{x}}_i^T\|}{l}) + \frac{r(r+t)}{2t} \mathsf{SNR}^{1+\alpha} + O(\mathsf{SNR}^{1+\alpha+\gamma}). \tag{54}$$

Combining (54) with (44) and noting that all the input vectors have identically distributed norms, we see that the input distribution satisfying (44) satisfies the second constraint (3) also. This completes the proof of the lemma. □

# Appendix 2

**Proof of Theorem 1**

For any $\alpha \in (0, 1]$ and $\gamma \in (0, \alpha)$, let

$$C(\mathsf{SNR}) \geq C^*(\mathsf{SNR}) = r\mathsf{SNR} - \frac{r(r+t)}{2t} \mathsf{SNR}^{1+\alpha} + O(\mathsf{SNR}^{1+\alpha+\gamma}).$$

This implies that there exists a probability distribution on $\mathbf{X}$ such that

$$I(\mathbf{X}; \mathbf{Y}) \geq C^*(\mathsf{SNR}).$$

From Lemma 1, we know that this distribution must satisfy the following constraints for all $i \in \{1, \ldots, t\}$:

$$\frac{t}{l} E\Big[\log(1 + \|\vec{\mathbf{x}}_i^T\|^2)\Big] \leq \frac{(r+t)}{2t} \mathsf{SNR}^{1+\alpha} + O(\mathsf{SNR}^{1+\alpha+\gamma}), \tag{55}$$

$$tE\Big[\log\Big(1 + \frac{\|\vec{\mathbf{x}}_i^T\|^2}{l}\Big)\Big] \geq \mathsf{SNR} - \frac{(r+t)}{t} \mathsf{SNR}^{1+\alpha} + O(\mathsf{SNR}^{1+\alpha+\gamma}). \tag{56}$$

Using these constraints, we establish a necessary condition on the coherence length. As the norms of the transmitted signals are identically distributed, it suffices to consider only one of them. Therefore, we will omit the subscript, $i$, and define random variable $\mathbf{b}$ as

$$\mathbf{b} \triangleq \frac{t\|\vec{\mathbf{x}}^T\|^2}{l\mathsf{SNR}}.$$

The two constraints become

$$\begin{aligned}
\frac{t}{l}E\Big[\log\Big(1+\frac{\mathbf{b}l\mathsf{SNR}}{t}\Big)\Big] &\le \frac{(r+t)}{2t}\mathsf{SNR}^{1+\alpha}+O(\mathsf{SNR}^{1+\alpha+\gamma}), && (57)\\
tE\Big[\log\Big(1+\frac{\mathbf{b}\mathsf{SNR}}{t}\Big)\Big] &\ge \mathsf{SNR}-\frac{(r+t)}{t}\mathsf{SNR}^{1+\alpha}+O(\mathsf{SNR}^{1+\alpha+\gamma}). && (58)
\end{aligned}$$

Moreover, as

$$\begin{aligned}
E[\|\vec{\mathbf{x}}^T\|^2] &= \frac{l\mathsf{SNR}}{t}, && (59)\\
\Rightarrow E[\mathbf{b}] &= 1. && (60)
\end{aligned}$$

Note that (58,60) do not depend on the coherence length, $l$, whereas (57) does. Also, the left hand side of (57) is a monotonically decreasing function of $l$. Thus, to find how large the coherence length necessarily needs to be, we need to find the distribution on $\mathbf{b}$ that minimizes the left hand side of (57) subject to the constraints (58,60). Using this distribution for $\mathbf{b}$, we can obtain the necessary condition on the coherence length from (57).
For any $\beta > 0$, we can express (58) as

$$\begin{aligned}
&\frac{\mathsf{SNR}}{t}-\frac{(r+t)}{t^2}\mathsf{SNR}^{1+\alpha}+O(\mathsf{SNR}^{1+\alpha+\gamma})\\
&\quad\le E\Big[\log\Big(1+\frac{\mathbf{b}\mathsf{SNR}}{t}\Big)\Big]\\
&\quad= \Pr(\mathbf{b}\ge t\mathsf{SNR}^{-\beta})E\Big[\log\Big(1+\frac{\mathbf{b}\mathsf{SNR}}{t}\Big)\Big|\mathbf{b}\ge t\mathsf{SNR}^{-\beta}\Big]\\
&\qquad\qquad+\Pr(\mathbf{b}< t\mathsf{SNR}^{-\beta})E\Big[\log\Big(1+\frac{\mathbf{b}\mathsf{SNR}}{t}\Big)\Big|\mathbf{b}< t\mathsf{SNR}^{-\beta}\Big]\\
&\quad\le \Pr(\mathbf{b}\ge t\mathsf{SNR}^{-\beta})E\Big[\log\Big(1+\frac{\mathbf{b}\mathsf{SNR}}{t}\Big)\Big|\mathbf{b}\ge t\mathsf{SNR}^{-\beta}\Big]+\Pr(\mathbf{b}< t\mathsf{SNR}^{-\beta})E\Big[\frac{\mathbf{b}\mathsf{SNR}}{t}\Big|\mathbf{b}< t\mathsf{SNR}^{-\beta}\Big]\\
&\quad= \frac{\mathsf{SNR}}{t}-\Pr(\mathbf{b}\ge t\mathsf{SNR}^{-\beta})E\Big[\frac{\mathbf{b}\mathsf{SNR}}{t}-\log\Big(1+\frac{\mathbf{b}\mathsf{SNR}}{t}\Big)\Big|\mathbf{b}\ge t\mathsf{SNR}^{-\beta}\Big].
\end{aligned}$$

Therefore,

$$\begin{aligned}
&\Pr(\mathbf{b}\ge t\mathsf{SNR}^{-\beta})E\Big[\frac{\mathbf{b}\mathsf{SNR}}{t}-\log\Big(1+\frac{\mathbf{b}\mathsf{SNR}}{t}\Big)\Big|\mathbf{b}\ge t\mathsf{SNR}^{-\beta}\Big] && (61)\\
&\qquad\qquad\le \frac{(r+t)}{t^2}\mathsf{SNR}^{1+\alpha}+O(\mathsf{SNR}^{1+\alpha+\gamma}).
\end{aligned}$$

When $\beta \ge 1$, $\mathbf{b} \ge t\mathsf{SNR}^{-\beta}$ implies $\frac{\mathbf{b}\mathsf{SNR}}{t} \gg 1$ which makes $\frac{\mathbf{b}\mathsf{SNR}}{t} \gg \log(1+\frac{\mathbf{b}\mathsf{SNR}}{t})$. Hence, $\forall\beta \ge 1$

$$\Pr(\mathbf{b}\ge t\mathsf{SNR}^{-\beta})E[\mathbf{b}|\mathbf{b}\ge t\mathsf{SNR}^{-\beta}] \le \frac{(r+t)}{t}\mathsf{SNR}^{\alpha}+O(\mathsf{SNR}^{\alpha+\gamma}) = o(1). \quad (62)$$

From Markov's inequality, $\forall\beta \ge 1$

$$\Pr(\mathbf{b}\ge t\mathsf{SNR}^{-\beta}) \le \frac{\mathsf{SNR}^{\beta}}{t} = o(1). \quad (63)$$

When $\beta < 1$, $\mathbf{b} < t\mathsf{SNR}^{-\beta}$ implies $\frac{\mathbf{b}\mathsf{SNR}}{t} \ll 1$. Hence, (61) can be expressed as

$$\begin{aligned}
&\frac{(r+t)}{t^2}\mathsf{SNR}^{1+\alpha} + O(\mathsf{SNR}^{1+\alpha+\gamma}) \\
&\quad \geq Pr(\mathbf{b} \geq t\mathsf{SNR}^{-\beta}) E\Big[\frac{\mathbf{b}\mathsf{SNR}}{t} - \log\Big(1 + \frac{\mathbf{b}\mathsf{SNR}}{t}\Big)\Big|\mathbf{b} \geq t\mathsf{SNR}^{-\beta}\Big] \\
&\quad \geq \Pr\Big(t\mathsf{SNR}^{-1} \geq \mathbf{b} \geq t\mathsf{SNR}^{-\beta}\Big) E\Big[\frac{\mathbf{b}\mathsf{SNR}}{t} - \log\Big(1 + \frac{\mathbf{b}\mathsf{SNR}}{t}\Big)\Big|t\mathsf{SNR}^{-1} \geq \mathbf{b} \geq t\mathsf{SNR}^{-\beta}\Big] \\
&\quad \geq \Pr\Big(t\mathsf{SNR}^{-1} \geq \mathbf{b} \geq t\mathsf{SNR}^{-\beta}\Big) E\Big[\frac{1}{2}\Big(\frac{\mathbf{b}\mathsf{SNR}}{t}\Big)^2 - \frac{1}{3}\Big(\frac{\mathbf{b}\mathsf{SNR}}{t}\Big)^3\Big|t\mathsf{SNR}^{-1} \geq \mathbf{b} \geq t\mathsf{SNR}^{-\beta}\Big] \\
&\quad \geq \Pr\Big(t\mathsf{SNR}^{-1} \geq \mathbf{b} \geq t\mathsf{SNR}^{-\beta}\Big)\Big[\frac{1}{2}\mathsf{SNR}^{2(1-\beta)} - \frac{1}{3}\mathsf{SNR}^{3(1-\beta)}\Big]
\end{aligned}$$

Thus, $\forall \beta \in (0,1)$

$$\Pr\Big(t\mathsf{SNR}^{-1} \geq \mathbf{b} \geq t\mathsf{SNR}^{-\beta}\Big) \leq \frac{2(r+t)}{t^2}\frac{\mathsf{SNR}^{2\beta-(1-\alpha)}}{1 - \frac{2}{3}\mathsf{SNR}^{1-\beta}} + o\Big(\frac{\mathsf{SNR}^{2\beta-(1-\alpha)}}{1 - \frac{2}{3}\mathsf{SNR}^{1-\beta}}\Big). \tag{64}$$

Let us divide the interval $[t\mathsf{SNR}^{-\beta}, t\mathsf{SNR}^{-1}]$, $\beta \in (0,1)$, into $K > 1$ finite intervals so that each interval is of length

$$\varepsilon = \frac{t(\mathsf{SNR}^{-1} - \mathsf{SNR}^{-\beta})}{K}.$$

Now, for any $\varepsilon > 0$

$$\begin{aligned}
&E\Big[\mathbf{b}|t\mathsf{SNR}^{-1} \geq \mathbf{b} \geq t\mathsf{SNR}^{-\beta}\Big]\Pr\Big(t\mathsf{SNR}^{-1} \geq \mathbf{b} \geq t\mathsf{SNR}^{-\beta}\Big) \\
&\quad = \sum_{i=1}^{K} E\Big[\mathbf{b}|t\mathsf{SNR}^{-(\beta+i\varepsilon)} \geq \mathbf{b} \geq t\mathsf{SNR}^{-[\beta+(i-1)\varepsilon]}\Big]\Pr\Big(t\mathsf{SNR}^{-(\beta+i\varepsilon)} \geq \mathbf{b} \geq t\mathsf{SNR}^{-[\beta+(i-1)\varepsilon]}\Big) \\
&\quad \leq t\sum_{i=1}^{K}\mathsf{SNR}^{-(\beta+i\varepsilon)}\Pr\Big(t\mathsf{SNR}^{-(\beta+i\varepsilon)} \geq \mathbf{b} \geq t\mathsf{SNR}^{-[\beta+(i-1)\varepsilon]}\Big) \\
&\quad \leq t\sum_{i=1}^{K}\mathsf{SNR}^{-(\beta+i\varepsilon)}\Pr\Big(t\mathsf{SNR}^{-1} \geq \mathbf{b} \geq t\mathsf{SNR}^{-[\beta+(i-1)\varepsilon]}\Big) \\
&\quad \leq \frac{2(r+t)}{t}\sum_{i=1}^{K}\mathsf{SNR}^{-(\beta+i\varepsilon)}\Big[\frac{\mathsf{SNR}^{[2(\beta+(i-1)\varepsilon)-(1-\alpha)]}}{1 - \frac{2}{3}\mathsf{SNR}^{[1-(\beta+(i-1)\varepsilon)]}} + o\Big(\frac{\mathsf{SNR}^{[2(\beta+(i-1)\varepsilon)-(1-\alpha)]}}{1 - \frac{2}{3}\mathsf{SNR}^{[1-(\beta+(i-1)\varepsilon)]}}\Big)\Big] \qquad (65) \\
&\quad = \frac{2(r+t)}{t}\sum_{i=1}^{K}\Big[\frac{\mathsf{SNR}^{[\beta-(1-\alpha)+(i-2)\varepsilon]}}{1 - \frac{2}{3}\mathsf{SNR}^{[1-(\beta+(i-1)\varepsilon)]}} + o\Big(\frac{\mathsf{SNR}^{[\beta-(1-\alpha)+(i-2)\varepsilon]}}{1 - \frac{2}{3}\mathsf{SNR}^{[1-(\beta+(i-1)\varepsilon)]}}\Big)\Big].
\end{aligned}$$

Equation (64) is used to obtain (65). Let $\beta = 1 - \alpha + 2\varepsilon$, where $\varepsilon \in (0, \frac{\alpha}{2})$. Thus,

$$\begin{aligned}
&E[\mathbf{b}|t\mathsf{SNR}^{-1} \geq \mathbf{b} \geq t\mathsf{SNR}^{-\beta}]\Pr(t\mathsf{SNR}^{-1} \geq \mathbf{b} \geq t\mathsf{SNR}^{-\beta}) \\
&\quad \leq \frac{2(r+t)}{t}\sum_{i=1}^{K}\Big[\frac{\mathsf{SNR}^{i\varepsilon}}{1 - \frac{2}{3}\mathsf{SNR}^{\alpha-(i+1)\varepsilon}} + o\Big(\frac{\mathsf{SNR}^{i\varepsilon}}{1 - \frac{2}{3}\mathsf{SNR}^{\alpha-(i+1)\varepsilon}}\Big)\Big].
\end{aligned}$$

Since,

$$\frac{\mathsf{SNR}^{i\varepsilon}}{1-\frac{2}{3}\mathsf{SNR}^{\alpha-(i+1)\varepsilon}} = o(1) \quad \forall i \in \{1,\ldots,K\},$$

we have

$$E\Big[\mathbf{b}|t\mathsf{SNR}^{-1} \geq \mathbf{b} \geq t\mathsf{SNR}^{-(1-\alpha+2\varepsilon)}\Big] \Pr\Big(t\mathsf{SNR}^{-1} \geq \mathbf{b} \geq t\mathsf{SNR}^{-(1-\alpha+2\varepsilon)}\Big) \leq o(1). \tag{66}$$

From (64), we have for $\beta = 1-\alpha+2\varepsilon$, where $\varepsilon \in (0, \frac{\alpha}{2})$,

$$\Pr\Big(t\mathsf{SNR}^{-1} \geq \mathbf{b} \geq t\mathsf{SNR}^{-\beta}\Big) \leq \frac{2(r+t)}{t^2}\frac{\mathsf{SNR}^{1-\alpha+4\varepsilon}}{1-\frac{2}{3}\mathsf{SNR}^{\alpha-2\varepsilon}} + o\Big(\frac{\mathsf{SNR}^{1-\alpha+4\varepsilon}}{1-\frac{2}{3}\mathsf{SNR}^{\alpha-2\varepsilon}}\Big) = o(1). \tag{67}$$

From (62,63,66,67), we know that for $0 < \epsilon < \alpha$

$$\begin{aligned} \Pr(\mathbf{b} \geq t\mathsf{SNR}^{-(1-\alpha+\epsilon)}) &= o(1), \\ \Pr(\mathbf{b} \geq t\mathsf{SNR}^{-(1-\alpha+\epsilon)})E[\mathbf{b}|\mathbf{b} \geq t\mathsf{SNR}^{-(1-\alpha+\epsilon)}] &= o(1), \end{aligned}$$

which implies

$$\begin{aligned} \Pr(\mathbf{b} \leq t\mathsf{SNR}^{-(1-\alpha+\epsilon)}) &= O(1), \\ \Pr(\mathbf{b} \leq t\mathsf{SNR}^{-(1-\alpha+\epsilon)})E[\mathbf{b}|\mathbf{b} \leq t\mathsf{SNR}^{-(1-\alpha+\epsilon)}] &= O(1). \end{aligned}$$

Hence, the distribution on $\mathbf{b}$ that minimizes $E\Big[\log\Big(1+\frac{\mathbf{b}l\mathsf{SNR}}{t}\Big)\Big]$, subject to the constraints (58,60), is the on-off distribution:

$$\mathbf{b} = \begin{cases} t\mathsf{SNR}^{-(1-\alpha+\epsilon)} & w.p. \;\; \eta \\ 0 & w.p. \;\; 1-\eta \end{cases}$$

where,

$$\eta = \frac{\mathsf{SNR}^{1-\alpha+\epsilon}}{t}.$$

Hence, with this on-off distribution on $\mathbf{b}$, (57) becomes

$$\frac{(r+t)}{2t}\mathsf{SNR}^{1+\alpha} + O(\mathsf{SNR}^{1+\alpha+\gamma}) \geq \frac{\mathsf{SNR}^{1-\alpha+\epsilon}}{l}\log\Big(1+l\mathsf{SNR}^{\alpha-\epsilon}\Big). \tag{68}$$

Now,

$$\begin{aligned} &\frac{\mathsf{SNR}^{1-\alpha+\epsilon}}{l}\log\Big(1+l\mathsf{SNR}^{\alpha-\epsilon}\Big)\Big|_{l=\frac{t^2}{(r+t)^2}\mathsf{SNR}^{-2\alpha+\epsilon}} \\ &= \frac{(r+t)^2}{t^2}\mathsf{SNR}^{1+\alpha}\log\Big(1+\frac{t^2}{(r+t)^2}\mathsf{SNR}^{-\alpha}\Big) \\ &\gg \frac{(r+t)}{2t}\mathsf{SNR}^{1+\alpha} + O(\mathsf{SNR}^{1+\alpha+\gamma}). \end{aligned}$$

Thus, with $l = \frac{t^2}{(r+t)^2}\mathsf{SNR}^{-2\alpha+\epsilon}$, (68) is not satisfied. However, the right hand side of (68) is a monotonically decreasing function of $l$. Hence, for the constraint in (68) to be met

$$l > \frac{t^2}{(r+t)^2}\mathsf{SNR}^{-2\alpha+\epsilon} \quad \forall \epsilon \in (0, \alpha).$$

Thus, we see that if an input distribution satisfies (55,56,59), then the coherence length must necessarily obey

$$l > \frac{t^2}{(r+t)^2}\mathsf{SNR}^{-2\alpha} \triangleq l_{\mathsf{min}}.$$

This completes the proof of the theorem. □

# Appendix 3

We compute the capacity of an i.i.d Rayleigh fading MIMO channel when CSI is unavailable at both the transmitter as well as the receiver. It is shown in [8] that increasing the number of transmit antennas beyond the coherence length does not increase capacity. Hence, from a capacity point of view, it suffices to use only one transmit antenna ($t = 1$). We will therefore consider the capacity of a single-input, multiple-output (SIMO) channel.
Let us pick on-off signaling to communicate over the channel. This signaling scheme is later proved to be optimal for the i.i.d Rayleigh fading MIMO channel. We specify the signaling as

$$\mathbf{x} = \begin{cases} \sqrt{A} & w.p.\ \omega \\ 0 & w.p.\ 1-\omega \end{cases} \tag{69}$$

where, $A \in \Re^+$ and $\omega = \frac{\mathsf{SNR}}{A}$. With this signaling, we have the following probability distributions

$$\begin{aligned} p_{\vec{\mathbf{y}}|\mathbf{x}=0}(\vec{y}) &= \frac{1}{\pi^r}\exp(-\|\vec{y}\|^2), \\ p_{\vec{\mathbf{y}}|\mathbf{x}=\sqrt{A}}(\vec{y}) &= \frac{1}{\pi(1+A)^r}\exp\Big(-\frac{\|\vec{y}\|^2}{1+A}\Big). \end{aligned}$$

The mutual information $I(\mathbf{x}, \vec{\mathbf{y}})$ can be written as $I(\mathbf{x}, \vec{\mathbf{y}}) = h(\vec{\mathbf{y}}) - h(\vec{\mathbf{y}}|\mathbf{x})$. Now,

$$\begin{aligned} & h(\vec{\mathbf{y}}) \\ &= -\int p_{\vec{\mathbf{y}}}(\vec{y})\log(p_{\vec{\mathbf{y}}}(\vec{y}))d\vec{y} \\ &= -\int \Big[(1-\omega)p_{\vec{\mathbf{y}}|\mathbf{x}=0}(\vec{y}) + \omega p_{\vec{\mathbf{y}}|\mathbf{x}=\sqrt{A}}(\vec{y})\Big]\log\Big[(1-\omega)p_{\vec{\mathbf{y}}|\mathbf{x}=0}(\vec{y}) + \omega p_{\vec{\mathbf{y}}|\mathbf{x}=\sqrt{A}}(\vec{y})\Big]d\vec{y} \\ &= -(1-\omega)\int p_{\vec{\mathbf{y}}|\mathbf{x}=0}\log\Big[(1-\omega)p_{\vec{\mathbf{y}}|\mathbf{x}=0}(\vec{y})(1+\frac{\omega}{1-\omega}\frac{p_{\vec{\mathbf{y}}|\mathbf{x}=\sqrt{A}}(\vec{y})}{p_{\vec{\mathbf{y}}|\mathbf{x}=0}(\vec{y})})\Big]d\vec{y} \\ & -\omega\int p_{\vec{\mathbf{y}}|\mathbf{x}=\sqrt{A}}\log\Big[(1-\omega)p_{\vec{\mathbf{y}}|\mathbf{x}=0}(\vec{y})(1+\frac{\omega}{1-\omega}\frac{p_{\vec{\mathbf{y}}|\mathbf{x}=\sqrt{A}}(\vec{y})}{p_{\vec{\mathbf{y}}|\mathbf{x}=0}(\vec{y})})\Big]d\vec{y} \end{aligned}$$

$$= -\log(1-\omega) - (1-\omega)\int p_{\vec{\mathbf{y}}|\mathbf{x}=0}(\vec{y})\log\left[1+\frac{\omega}{1-\omega}\frac{p_{\vec{\mathbf{y}}|\mathbf{x}=\sqrt{A}}(\vec{y})}{p_{\vec{\mathbf{y}}|\mathbf{x}=0}(\vec{y})}\right]d\vec{y}$$

$$-\omega\int p_{\vec{\mathbf{y}}|\mathbf{x}=\sqrt{A}}(\vec{y})\log\left[1+\frac{\omega}{1-\omega}\frac{p_{\vec{\mathbf{y}}|\mathbf{x}=\sqrt{A}}(\vec{y})}{p_{\vec{\mathbf{y}}|\mathbf{x}=0}(\vec{y})}\right]d\vec{y} + h(\vec{\mathbf{y}}|\mathbf{x}) + \omega D(p_{\vec{\mathbf{y}}|\mathbf{x}=\sqrt{A}}||p_{\vec{\mathbf{y}}|\mathbf{x}=0})$$

The divergence $D(p_{\vec{\mathbf{y}}|\mathbf{x}=\sqrt{A}}||p_{\vec{\mathbf{y}}|\mathbf{x}=0})$ is the divergence between two Gaussian random vectors and is therefore

$$D(p_{\vec{\mathbf{y}}|\mathbf{x}=\sqrt{A}}||p_{\vec{\mathbf{y}}|\mathbf{x}=0}) = r(A - \log(1+A))$$

The expression for the mutual information becomes

$$I(\mathbf{x},\vec{\mathbf{y}}) = r\mathsf{SNR} - r\mathsf{SNR}\frac{\log(1+A)}{A} - \log(1-\frac{\mathsf{SNR}}{A}) - I(\mathsf{SNR},A)$$

where,

$$I(\mathsf{SNR},A) = I_1(\mathsf{SNR},A) + I_2(\mathsf{SNR},A), \tag{70}$$

$$I_1(\mathsf{SNR},A) = (1-\omega)\int p_{\vec{\mathbf{y}}|\mathbf{x}=0}(\vec{y})\log\left[1+\frac{\omega}{1-\omega}\frac{p_{\vec{\mathbf{y}}|\mathbf{x}=\sqrt{A}}(\vec{y})}{p_{\vec{\mathbf{y}}|\mathbf{x}=0}(\vec{y})}\right]d\vec{y}$$

$$I_2(\mathsf{SNR},A) = \omega\int p_{\vec{\mathbf{y}}|\mathbf{x}=\sqrt{A}}(\vec{y})\log\left[1+\frac{\omega}{1-\omega}\frac{p_{\vec{\mathbf{y}}|\mathbf{x}=\sqrt{A}}(\vec{y})}{p_{\vec{\mathbf{y}}|\mathbf{x}=0}(\vec{y})}\right]d\vec{y}$$

At low $\mathsf{SNR}$, $A$ takes very high values. therefore the mutual information can be written as

$$I(\mathbf{x},\vec{\mathbf{y}}) = r\mathsf{SNR} - r\mathsf{SNR}\frac{\log(1+A)}{A} - I(\mathsf{SNR},A) \tag{71}$$

We will now compute $I_1(\mathsf{SNR},A)$ and $I_2(\mathsf{SNR},A)$. Let us define $\varsigma^*$ to be such that

$$\frac{\mathsf{SNR}}{A(1+A)^r}\exp\left(\frac{A\varsigma^*}{1+A}\right) = 1.$$

Note that

$$\frac{\varsigma^*}{1+A} = \frac{\log(A)}{A} + r\frac{\log(1+A)}{A} + \frac{\log(\frac{1}{\mathsf{SNR}})}{A}.$$

Thus,

$$\lim_{A\to\infty}\frac{\varsigma^*}{1+A} = 0.$$

We will use this in future derivations.

**Computing** $I_1(\mathsf{SNR},A)$:

$$I_1(\mathsf{SNR},A) = \frac{1}{\pi^r}(1-\frac{\mathsf{SNR}}{A})\int\exp(-\|\vec{y}\|^2)\log[1+\frac{\mathsf{SNR}}{(A-\mathsf{SNR})(1+A)^r}\exp(\frac{A}{1+A}\|\vec{y}\|^2)]d\vec{y}$$

Converting to spherical coordinates in $2r$ dimensions, we have for large $A$:

$$\begin{aligned}
&I_1(\mathsf{SNR}, A) \\
&= \frac{1}{(r-1)!}\int_0^\infty \varsigma^{r-1}\exp(-\varsigma)\log\Big[1+\frac{\mathsf{SNR}}{A(1+A)^r}\exp\Big(\frac{A\varsigma}{1+A}\Big)\Big]d\varsigma + o(\mathsf{SNR}^2) \\
&= \frac{\exp(-\varsigma^*)}{(r-1)!}\int_0^\infty \varsigma^{r-1}\exp(-(\varsigma-\varsigma^*))\log\Big[1+\exp\Big(\frac{A(\varsigma-\varsigma^*)}{1+A}\Big)\Big]d\varsigma + o(\mathsf{SNR}^2) \\
&= \frac{\exp(-\varsigma^*)}{(r-1)!}\int_{-\varsigma^*}^\infty (\varsigma+\varsigma^*)^{r-1}\exp(-\varsigma)\log\Big[1+\exp\Big(\frac{A\varsigma}{1+A}\Big)\Big]d\varsigma + o(\mathsf{SNR}^2) \\
&= I_{11}(\mathsf{SNR}, A) + I_{12}(\mathsf{SNR}, A) + o(\mathsf{SNR}^2),
\end{aligned} \tag{72}$$

where,

$$\begin{aligned}
I_{11}(\mathsf{SNR}, A) &= \frac{\exp(-\varsigma^*)}{(r-1)!}\int_{-\varsigma^*}^0 (\varsigma+\varsigma^*)^{r-1}\exp(-\varsigma)\log\Big[1+\exp\Big(\frac{A\varsigma}{1+A}\Big)\Big]d\varsigma \\
I_{12}(\mathsf{SNR}, A) &= \frac{\exp(-\varsigma^*)}{(r-1)!}\int_0^\infty (\varsigma+\varsigma^*)^{r-1}\exp(-\varsigma)\log\Big[1+\exp\Big(\frac{A\varsigma}{1+A}\Big)\Big]d\varsigma
\end{aligned}$$

**Computing** $I_{11}(\mathsf{SNR}, A)$:

$$\begin{aligned}
&I_{11}(\mathsf{SNR}, A) \\
&= \frac{\exp(-\varsigma^*)}{(r-1)!}\int_{-\varsigma^*}^0 (\varsigma+\varsigma^*)^{r-1}\exp(-\varsigma)\log\Big[1+\exp\Big(\frac{A\varsigma}{1+A}\Big)\Big]d\varsigma \\
&\leq \frac{\exp(-\varsigma^*)}{(r-1)!}\int_{-\varsigma^*}^0 (\varsigma+\varsigma^*)^{r-1}\exp(-\varsigma)\exp\Big(\frac{A\varsigma}{1+A}\Big)d\varsigma \\
&= \frac{\exp(-\varsigma^*)}{(r-1)!}\int_{-\varsigma^*}^0 (\varsigma+\varsigma^*)^{r-1}\exp(-\frac{\varsigma}{1+A})d\varsigma \\
&= \frac{(1+A)^r}{(r-1)!}\exp(-\frac{A\varsigma^*}{1+A})\int_0^{\frac{\varsigma^*}{1+A}} \varsigma^{r-1}\exp(-\varsigma)d\varsigma \\
&= \frac{(1+A)^r}{(r-1)!}\exp(-\frac{A\varsigma^*}{1+A})\Big[\Gamma(r)-\Gamma(r,\frac{\varsigma^*}{1+A})\Big] \\
&= I_{11}^U(\mathsf{SNR}, A),
\end{aligned}$$

where,

$$I_{11}^U(\mathsf{SNR}, A) = \frac{\mathsf{SNR}}{A} - (1+A)^r\Big[\frac{\mathsf{SNR}}{A(1+A)^r}\Big]^{1+\frac{1}{A}}\Big[\sum_{j=0}^{r-1}\frac{[\frac{\varsigma^*}{1+A}]^j}{j!}\Big]$$

Moreover,

$$\begin{aligned}
&I_{11}(\mathsf{SNR}, A) \\
&= \frac{\exp(-\varsigma^*)}{(r-1)!}\int_{-\varsigma^*}^0 (\varsigma+\varsigma^*)^{r-1}\exp(-\varsigma)\log\Big[1+\exp\Big(\frac{A\varsigma}{1+A}\Big)\Big]d\varsigma
\end{aligned}$$

$$
\begin{aligned}
&\geq I_{11}^U(\mathsf{SNR},A) - \frac{\exp(-\varsigma^*)}{2(r-1)!}\int_{-\varsigma^*}^{0}(\varsigma+\varsigma^*)^{r-1}\exp(-\varsigma)\exp\Big(\frac{2A\varsigma}{1+A}\Big)d\varsigma \\
&= I_{11}^U(\mathsf{SNR},A) - \frac{\exp(-\varsigma^*)}{2(r-1)!}\int_{-\varsigma^*}^{0}(\varsigma+\varsigma^*)^{r-1}\exp\Big(\frac{A-1}{A+1}\varsigma\Big)d\varsigma \\
&= I_{11}^U(\mathsf{SNR},A) - \frac{\exp(-\varsigma^*)}{2(r-1)!}\int_{0}^{\varsigma^*}\varsigma^{r-1}\exp\Big(\frac{A-1}{A+1}(\varsigma-\varsigma^*)\Big)d\varsigma \\
&= I_{11}^U(\mathsf{SNR},A) - \frac{1}{2(r-1)!}\exp\Big(-\frac{2A\varsigma^*}{A+1}\Big)\Big[\frac{A+1}{A-1}\Big]^r\int_{0}^{(\frac{A-1}{A+1})\varsigma^*}\varsigma^{r-1}\exp(\varsigma)d\varsigma \\
&= I_{11}^U(\mathsf{SNR},A) - \frac{(-1)^{r-1}}{2(r-1)!}\exp\Big(-\frac{2A\varsigma^*}{A+1}\Big)\Big[\frac{A+1}{A-1}\Big]^r\Big[\Gamma(r,-\frac{A-1}{A+1}\varsigma^*)-\Gamma(r)\Big] \\
&= I_{11}^U(\mathsf{SNR},A) - \frac{(-1)^{r-1}}{2}\Big[\frac{A+1}{A-1}\Big]^r\Big[\exp(-\varsigma^*)\sum_{j=0}^{r-1}\frac{[-\frac{A-1}{A+1}\varsigma^*]^j}{j!} - \exp\Big(-\frac{2A\varsigma^*}{A+1}\Big)\Big] \\
&= I_{11}^U(\mathsf{SNR},A) - \frac{(-1)^{r-1}}{2}\Big[\frac{A+1}{A-1}\Big]^r\Big[\frac{\mathsf{SNR}}{A(1+A)^r}\Big]^{1+\frac{1}{1+A}}\Big[\sum_{j=0}^{r-1}\frac{[-\frac{A-1}{A+1}\varsigma^*]^j}{j!}\Big] + o(\mathsf{SNR}^2)
\end{aligned}
$$

Let

$$
I_{11}^L(\mathsf{SNR},A) = \frac{(-1)^{r-1}}{2}\Big[\frac{A+1}{A-1}\Big]^r\Big[\frac{\mathsf{SNR}}{A(1+A)^r}\Big]^{1+\frac{1}{1+A}}\Big[\sum_{j=0}^{r-1}\frac{[-\frac{A-1}{A+1}\varsigma^*]^j}{j!}\Big]
$$

Thus, we have

$$
I_{11}^U(\mathsf{SNR},A) - I_{11}^L(\mathsf{SNR},A) + o(\mathsf{SNR}^2) \leq I_{11}(\mathsf{SNR},A) \leq I_{11}^U(\mathsf{SNR},A)
$$

Since $\frac{\varsigma^*}{1+A} \to 0$ as $A \to \infty$, we have

$$
\begin{aligned}
&\lim_{A\to\infty} I_{11}^U(\mathsf{SNR},A) = 0, \\
&\lim_{A\to\infty} I_{11}^L(\mathsf{SNR},A) = 0, \\
\Rightarrow &\lim_{A\to\infty} I_{11}(\mathsf{SNR},A) = 0.
\end{aligned} \tag{73}
$$

**Computing** $I_{12}(\mathsf{SNR},A)$:

$$
\begin{aligned}
&I_{12}(\mathsf{SNR},A) \\
&\quad= \frac{\exp(-\varsigma^*)}{(r-1)!}\int_0^\infty(\varsigma+\varsigma^*)^{r-1}\exp(-\varsigma)\log\Big[1+\exp\Big(\frac{A\varsigma}{1+A}\Big)\Big]d\varsigma \\
&\quad= \frac{\exp(-\varsigma^*)}{(r-1)!}\int_0^\infty(\varsigma+\varsigma^*)^{r-1}\exp(-\varsigma)\Big[\frac{A\varsigma}{1+A}+\log\Big[1+\exp\Big(-\frac{A\varsigma}{1+A}\Big)\Big]\Big]d\varsigma \\
&\quad= I_{12}^1(\mathsf{SNR},A) + I_{12}^2(\mathsf{SNR},A),
\end{aligned} \tag{74}
$$

where,

$$
I_{12}^1(\mathsf{SNR},A) = \frac{\exp(-\varsigma^*)}{(r-1)!}\Big[\frac{A}{1+A}\Big]\int_0^\infty\varsigma(\varsigma+\varsigma^*)^{r-1}\exp(-\varsigma)d\varsigma,
$$

$$
I_{12}^2(\mathsf{SNR},A) = \frac{\exp(-\varsigma^*)}{(r-1)!}\int_0^\infty(\varsigma+\varsigma^*)^{r-1}\exp(-\varsigma)\log\Big[1+\exp\Big(-\frac{A\varsigma}{1+A}\Big)\Big]d\varsigma.
$$

Now,

$$
\begin{aligned}
&I_{12}^{1}(\mathsf{SNR}, A)\\
&\quad = \frac{\exp(-\varsigma^*)}{(r-1)!}\Big[\frac{A}{1+A}\Big]\int_0^\infty \varsigma(\varsigma+\varsigma^*)^{r-1}\exp(-\varsigma)d\varsigma,\\
&\quad = \frac{1}{(r-1)!}\Big[\frac{A}{1+A}\Big]\int_{\varsigma^*}^\infty (\varsigma-\varsigma^*)\varsigma^{r-1}\exp(-\varsigma)d\varsigma,\\
&\quad = \frac{1}{(r-1)!}\Big[\frac{A}{1+A}\Big]\Big[\Gamma(r+1,\varsigma^*)-\varsigma^*\Gamma(r,\varsigma^*)\Big]\\
&\quad = \frac{1}{(r-1)!}\Big[\frac{A}{1+A}\Big]\Big[\Gamma(r,\varsigma^*)(r-\varsigma^*)+(\varsigma^*)^r\exp(-\varsigma^*)\Big]\\
&\quad = \exp(-\varsigma^*)\Big[\frac{A}{1+A}\Big]\Big[\frac{r-\varsigma^*}{(r-1)!}\Big[\sum_{j=0}^{r-1}\frac{(\varsigma^*)^j}{j!}\Big]+\frac{(\varsigma^*)^r}{(r-1)!}\Big]\\
&\quad = \Big[\frac{\mathsf{SNR}}{A(1+A)^r}\Big]^{1+\frac{1}{A}}\Big[\frac{A}{1+A}\Big]\Big[\frac{r-\varsigma^*}{(r-1)!}\Big[\sum_{j=0}^{r-1}\frac{(\varsigma^*)^j}{j!}\Big]+\frac{(\varsigma^*)^r}{(r-1)!}\Big]\\
&\quad = I_{12}^{1A}(\mathsf{SNR}, A)\mathsf{SNR}^{1+\frac{1}{A}},
\end{aligned}
$$

where,

$$
I_{12}^{1A}(\mathsf{SNR}, A) = \Big[\frac{1}{A(1+A)^r}\Big]^{1+\frac{1}{A}}\Big[\frac{A}{1+A}\Big]\Big[\frac{r-\varsigma^*}{(r-1)!}\Big[\sum_{j=0}^{r-1}\frac{(\varsigma^*)^j}{j!}\Big]+\frac{(\varsigma^*)^r}{(r-1)!}\Big]
$$

Since $\frac{\varsigma^*}{1+A}\to 0$ as $A\to\infty$, we have

$$
\lim_{A\to\infty} I_{12}^{1A}(\mathsf{SNR}, A) = 0.
$$

Thus,

$$
\lim_{A\to\infty} I_{12}^{1}(\mathsf{SNR}, A) = 0. \tag{75}
$$

We will now compute $I_{12}^{2}(\mathsf{SNR}, A)$.

$$
\begin{aligned}
&I_{12}^{2}(\mathsf{SNR}, A)\\
&\quad = \frac{\exp(-\varsigma^*)}{(r-1)!}\int 0^\infty(\varsigma+\varsigma^*)^{r-1}\exp(-\varsigma)\log\Big[1+\exp\Big(-\frac{A\varsigma}{1+A}\Big)\Big]d\varsigma\\
&\quad \le \frac{\exp(-\varsigma^*)}{(r-1)!}\int_0^\infty(\varsigma+\varsigma^*)^{r-1}\exp(-\varsigma)\exp\Big(-\frac{A\varsigma}{1+A}\Big)d\varsigma\\
&\quad = \frac{\exp(-\varsigma^*)}{(r-1)!}\int_0^\infty(\varsigma+\varsigma^*)^{r-1}\exp(-\Big(\frac{1+2A}{1+A}\Big)\varsigma)d\varsigma\\
&\quad = \frac{1}{(r-1)!}\Big[\frac{1+A}{1+2A}\Big]^r\exp\Big(\frac{A\varsigma}{1+A}\Big)\int_{(\frac{1+2A}{1+A})\varsigma^*}^\infty \varsigma^{r-1}\exp(-\varsigma)d\varsigma
\end{aligned}
$$

$$= \frac{1}{(r-1)!}\Big[\frac{1+A}{1+2A}\Big]^r \exp\Big(\frac{A\varsigma}{1+A}\Big)\Gamma(r, (\frac{1+2A}{1+A})\varsigma^*)$$
$$= \Big[\frac{1+A}{1+2A}\Big]^r \exp(-\varsigma^*)\sum_{j=0}^{r-1}\frac{[(\frac{1+2A}{1+A})\varsigma^*]^j}{j!}$$
$$= I_{12}^{2U}(\mathsf{SNR}, A),$$

where,

$$I_{12}^{2U}(\mathsf{SNR}, A) = \Big[\frac{\mathsf{SNR}}{A(1+A)^r}\Big]^{1+\frac{1}{A}}\Big[\frac{1+A}{1+2A}\Big]^r\Big[\sum_{j=0}^{r-1}\frac{[(\frac{1+2A}{1+A})\varsigma^*]^j}{j!}\Big]$$

Moreover,

$$I_{12}^{2}(\mathsf{SNR}, A)$$
$$\geq I_{12}^{2U}(\mathsf{SNR}, A) - \frac{\exp(-\varsigma^*)}{2(r-1)!}\int_0^\infty (\varsigma+\varsigma^*)^{r-1}\exp(-\varsigma)\exp\Big(-\frac{2A\varsigma}{1+A}\Big)d\varsigma$$
$$= I_{12}^{2U}(\mathsf{SNR}, A) - \frac{1}{2(r-1)!}\exp\Big(\frac{2A\varsigma}{1+A}\Big)\int_{\varsigma^*}^\infty \varsigma^{r-1}\exp(-\Big(\frac{1+3A}{1+A}\Big)\varsigma)d\varsigma$$
$$= I_{12}^{2U}(\mathsf{SNR}, A) - \frac{1}{2(r-1)!}\Big[\frac{1+A}{1+3A}\Big]^r\exp\Big(\frac{2A\varsigma}{1+A}\Big)\Gamma(r, (\frac{1+3A}{1+A})\varsigma^*)$$
$$= I_{12}^{2U}(\mathsf{SNR}, A) - \frac{\exp(-\varsigma^*)}{2}\Big[\frac{1+A}{1+3A}\Big]^r\Big[\sum_{j=0}^{r-1}\frac{[(\frac{1+3A}{1+A})\varsigma^*]^j}{j!}\Big]$$
$$= I_{12}^{2U}(\mathsf{SNR}, A) - \frac{1}{2}I_{12}^{2L}(\mathsf{SNR}, A).$$

where,

$$I_{12}^{2L}(\mathsf{SNR}, A) = \Big[\frac{\mathsf{SNR}}{A(1+A)^r)}\Big]^{1+\frac{1}{A}}\Big[\frac{1+A}{1+3A}\Big]^r\Big[\sum_{j=0}^{r-1}\frac{[(\frac{1+3A}{1+A})\varsigma^*]^j}{j!}\Big].$$

Since $\frac{\varsigma*}{1+A} \to 0$ as $A \to \infty$,

$$\lim_{A\to\infty} I_{12}^{2L}(\mathsf{SNR}, A) = 0,$$
$$\lim_{A\to\infty} I_{12}^{2U}(\mathsf{SNR}, A) = 0,$$
$$\Rightarrow \lim_{A\to\infty} I_{12}^{2}(\mathsf{SNR}, A) = 0. \quad (76)$$

Substituting (75) and (76) in (74), we have

$$\lim_{A\to\infty} I_{12}(\mathsf{SNR}, A) = 0. \quad (77)$$

Substituting (73) and (77) in (72), we obtain

$$I_1(\mathsf{SNR}, A) = o(\mathsf{SNR}^2). \quad (78)$$

**Computing** $I_2(\mathsf{SNR}, A)$:

$$I_2(\mathsf{SNR}, A) = \frac{\mathsf{SNR}}{\pi^r A(1+A)^r} \int \exp(-\frac{\|\vec{y}\|^2}{1+A}) \log[1 + \frac{\mathsf{SNR}}{(A-\mathsf{SNR})(1+A)^r} \exp(\frac{A}{1+A}\|\vec{y}\|^2)]d\vec{y}$$

Converting to spherical coordinates in $2r$ dimensions, we have for large $A$:

$$\begin{aligned} & I_2(\mathsf{SNR}, A) \\ & = \frac{\mathsf{SNR}}{A(A+1)^r(r-1)!} \int_0^{\infty} \varsigma^{r-1} \exp(-\frac{\varsigma}{1+A}) \log[1 + \frac{\mathsf{SNR}}{A(1+A)^r} \exp(\frac{A\varsigma}{1+A})]d\varsigma \\ & = I_{21}(\mathsf{SNR}, A) + I_{22}(\mathsf{SNR}, A) + o(\mathsf{SNR}^2), \end{aligned} \tag{79}$$

where

$$I_{21}(\mathsf{SNR}, A) = \frac{\mathsf{SNR}}{A(A+1)^r(r-1)!} \int_0^{\varsigma^*} \varsigma^{r-1} \exp(-\frac{\varsigma}{1+A}) \log[1 + \frac{\mathsf{SNR}}{A(1+A)^r} \exp(\frac{A\varsigma}{1+A})]d\varsigma,$$

$$I_{22}(\mathsf{SNR}, A) = \frac{\mathsf{SNR}}{A(A+1)^r(r-1)!} \int_{\varsigma^*}^{\infty} \varsigma^{r-1} \exp(-\frac{\varsigma}{1+A}) \log[1 + \frac{\mathsf{SNR}}{A(1+A)^r} \exp(\frac{A\varsigma}{1+A})]d\varsigma.$$

**Computing** $I_{21}(\mathsf{SNR}, A)$:

$$\begin{aligned} & I_{21}(\mathsf{SNR}, A) \\ & = \frac{\mathsf{SNR}}{A(A+1)^r(r-1)!} \int_0^{\varsigma^*} \varsigma^{r-1} \exp(-\frac{\varsigma}{1+A}) \log\left[1 + \frac{\mathsf{SNR}}{A(1+A)^r} \exp\left(\frac{A\varsigma}{1+A}\right)\right]d\varsigma \\ & \leq \frac{\mathsf{SNR}}{A(A+1)^r(r-1)!} \int_0^{\varsigma^*} \varsigma^{r-1} \exp\left(-\frac{\varsigma}{1+A}\right) \frac{\mathsf{SNR}}{A(1+A)^r} \exp\left(\frac{A\varsigma}{1+A}\right)d\varsigma \\ & = \left[\frac{\mathsf{SNR}}{A(1+A)^r}\right]^2 \frac{1}{(r-1)!} \int_0^{\varsigma^*} \varsigma^{r-1} \exp\left(\frac{A-1}{A+1}\varsigma\right)d\varsigma \\ & = \left[\frac{\mathsf{SNR}}{A(1+A)^r}\right]^2 \left[\frac{A+1}{A-1}\right]^r \frac{1}{(r-1)!} \int_0^{(\frac{A-1}{A+1})\varsigma^*} \varsigma^{r-1} \exp(\varsigma)d\varsigma \\ & = \left[\frac{\mathsf{SNR}}{A(1+A)^r}\right]^2 \left[\frac{A+1}{A-1}\right]^r \frac{(-1)^{r-1}}{(r-1)!} \left[\Gamma(r, -\left(\frac{A-1}{A+1}\right)\varsigma^*) - \Gamma(r)\right] \\ & = \left[\frac{\mathsf{SNR}}{A(1+A)^r}\right]^2 \left[\frac{A+1}{A-1}\right]^r \frac{(-1)^{r-1}}{(r-1)!} \left[(r-1)! \exp(\frac{A-1}{A+1})\varsigma^*) \sum_{j=0}^{r-1} \frac{-(\frac{A-1}{A+1})\varsigma^*}{j!} - (r-1)!\right] \\ & = \left[\frac{\mathsf{SNR}}{A(1+A)^r}\right]^2 \left[\frac{A+1}{A-1}\right]^r \frac{(-1)^{r-1}}{(r-1)!} \left[(r-1)! \left[\frac{\mathsf{SNR}}{A(1+A)^r}\right]^{\frac{1}{A}-1} \sum_{j=0}^{r-1} \frac{-(\frac{A-1}{A+1})\varsigma^*}{j!} - (r-1)!\right] \\ & = \left[I_{21}^U(\mathsf{SNR}, A)\right]\mathsf{SNR}^{1+\frac{1}{A}} + o(\mathsf{SNR}^2) \end{aligned}$$

where,

$$I_{21}^U(\mathsf{SNR}, A) = \frac{(-1)^r}{[A(1+A)^r]^{1+\frac{1}{A}}} \left[\frac{A+1}{A-1}\right]^r \sum_{j=0}^{r-1} \frac{[-(\frac{A-1}{A+1})\varsigma^*]^j}{j!}$$

Now,

$$
\begin{aligned}
&I_{21}(\mathsf{SNR}, A)\\
&= \frac{\mathsf{SNR}}{A(A+1)^r(r-1)!}\int_0^{\varsigma^*} \varsigma^{r-1}\exp\Big(-\frac{\varsigma}{1+A}\Big)\log[1+\frac{\mathsf{SNR}}{A(1+A)^r}\exp\Big(\frac{A\varsigma}{1+A}\Big)]d\varsigma\\
&\geq \frac{\mathsf{SNR}}{A(A+1)^r(r-1)!}\int_0^{\varsigma^*} \varsigma^{r-1}\exp\Big(-\frac{\varsigma}{1+A}\Big)\frac{\mathsf{SNR}}{A(1+A)^r}\exp\Big(\frac{A\varsigma}{1+A}\Big)d\varsigma\\
&-\frac{\mathsf{SNR}}{2A(A+1)^r(r-1)!}\int_0^{\varsigma^*} \varsigma^{r-1}\exp\Big(-\frac{\varsigma}{1+A}\Big)\Big[\frac{\mathsf{SNR}}{A(1+A)^r}\exp\Big(\frac{A\varsigma}{1+A}\Big)\Big]^2 d\varsigma\\
&= \Big[I_{21}^U(\mathsf{SNR}, A)\Big]\mathsf{SNR}^{1+\frac{1}{A}} + o(\mathsf{SNR}^2)\\
&-\frac{1}{2}\Big[\frac{\mathsf{SNR}}{A(1+A)^r}\Big]^3\frac{1}{(r-1)!}\int_0^{\varsigma^*}\varsigma^{r-1}\exp\Big(\frac{2A-1}{1+A}\varsigma\Big)d\varsigma\\
&= \Big[I_{21}^U(\mathsf{SNR}, A)\Big]\mathsf{SNR}^{1+\frac{1}{A}} + o(\mathsf{SNR}^2)\\
&-\frac{1}{2}\Big[\frac{\mathsf{SNR}}{A(1+A)^r}\Big]^3\frac{(-1)^r}{(r-1)!}\Big[\frac{A+1}{2A-1}\Big]^r\Big[\Gamma(r,-(\frac{2A-1}{1+A})\varsigma^*)-\Gamma(r)\Big]\\
&= \Big[I_{21}^U(\mathsf{SNR}, A)\Big]\mathsf{SNR}^{1+\frac{1}{A}} + o(\mathsf{SNR}^2)\\
&-\frac{1}{2}\Big[\frac{\mathsf{SNR}}{A(1+A)^r}\Big]^3\frac{(-1)^r}{(r-1)!}\Big[\frac{A+1}{2A-1}\Big]^r\Big[(r-1)!\Big[\frac{\mathsf{SNR}}{A(1+A)^r}\Big]^{\frac{1}{A}-2}\sum_{j=0}^{r-1}\frac{[-(\frac{2A-1}{A+1})\varsigma^*]^j}{j!}-(r-1)!\Big]\\
&= \Big[I_{21}^U(\mathsf{SNR}, A) - \frac{1}{2}I_{21}^L(\mathsf{SNR}, A)\Big]\mathsf{SNR}^{1+\frac{1}{A}} + o(\mathsf{SNR}^2)
\end{aligned}
$$

where,

$$
I_{21}^L(\mathsf{SNR}, A) = \frac{(-1)^r}{[A(1+A)^r]^{1+\frac{1}{A}}}\Big[\frac{A+1}{2A-1}\Big]^r\sum_{j=0}^{r-1}\frac{[-(\frac{2A-1}{A+1})\varsigma^*]^j}{j!}
$$

Combining the lower and upper bounds, we have

$$
\begin{aligned}
&\Big[I_{21}^U(\mathsf{SNR}, A) - \frac{1}{2}I_{21}^L(\mathsf{SNR}, A)\Big]\mathsf{SNR}^{1+\frac{1}{A}} + o(\mathsf{SNR}^2)\\
&\qquad\qquad \leq I_{21}(\mathsf{SNR}, A) \;\; \leq \;\; \Big[I_{21}^U(\mathsf{SNR}, A)\Big]\mathsf{SNR}^{1+\frac{1}{A}} + o(\mathsf{SNR}^2).
\end{aligned}
$$

At low $\mathsf{SNR}$, $\frac{\varsigma^*}{1+A} \to 0$ as $A \to \infty$. Thus, we have

$$
\begin{aligned}
\lim_{A\to\infty} I_{21}^U(\mathsf{SNR}, A) = 0\\
\lim_{A\to\infty} I_{21}^L(\mathsf{SNR}, A) = 0
\end{aligned}
$$

Therefore,

$$
I_{21}(\mathsf{SNR}, A) = o(\mathsf{SNR}^2). \tag{80}
$$

**Computing** $I_{22}(\mathsf{SNR}, A)$:

$$
\begin{aligned}
&I_{22}(\mathsf{SNR}, A)\\
&\quad= \frac{\mathsf{SNR}}{A(A+1)^r(r-1)!}\int_{\varsigma^*}^{\infty}\varsigma^{r-1}\exp\Big(-\frac{\varsigma}{1+A}\Big)\log\Big[1+\frac{\mathsf{SNR}}{A(1+A)^r}\exp\Big(\frac{A\varsigma}{1+A}\Big)\Big]d\varsigma\\
&\quad= \frac{1}{(r-1)!}\Big[\frac{\mathsf{SNR}}{A(1+A)^r}\Big]^{1+\frac{1}{A}}\int_{\varsigma^*}^{\infty}\varsigma^{r-1}\exp\Big(-\frac{(\varsigma-\varsigma^*)}{1+A}\Big)\log\Big[1+\exp\Big(\frac{A(\varsigma-\varsigma^*)}{1+A}\Big)\Big]d\varsigma\\
&\quad= \frac{1}{(r-1)!}\Big[\frac{\mathsf{SNR}}{A(1+A)^r}\Big]^{1+\frac{1}{A}}\int_{0}^{\infty}(\varsigma+\varsigma^*)^{r-1}\exp\Big(-\frac{\varsigma}{1+A}\Big)\log\Big[1+\exp\Big(\frac{A\varsigma}{1+A}\Big)\Big]d\varsigma\\
&\quad= \frac{1}{(r-1)!}\Big[\frac{\mathsf{SNR}}{A(1+A)^r}\Big]^{1+\frac{1}{A}}\int_{0}^{\infty}(\varsigma+\varsigma^*)^{r-1}\exp\Big(-\frac{\varsigma}{1+A}\Big)\Big[\frac{A\varsigma}{1+A}+\log\Big[1+\exp\Big(-\frac{A\varsigma}{1+A}\Big)\Big]\Big]d\varsigma\\
&\quad= \frac{1}{(r-1)!}\Big[\frac{1}{A(1+A)^r}\Big]^{\frac{1}{A}}\mathsf{SNR}^{1+\frac{1}{A}}\Big[I_{22}^1(\mathsf{SNR}, A)+I_{22}^2(\mathsf{SNR}, A)\Big] \qquad (81)
\end{aligned}
$$

where,

$$
\begin{aligned}
I_{22}^1(\mathsf{SNR}, A) &= \frac{1}{(1+A)^{r+1}}\int_0^{\infty}\varsigma(\varsigma+\varsigma^*)^{r-1}\exp\Big(-\frac{\varsigma}{1+A}\Big)d\varsigma,\\
I_{22}^2(\mathsf{SNR}, A) &= \frac{1}{A(1+A)^r}\int_0^{\infty}(\varsigma+\varsigma^*)^{r-1}\exp\Big(-\frac{\varsigma}{1+A}\Big)\log\Big[1+\exp\Big(-\frac{A\varsigma}{1+A}\Big)\Big]d\varsigma.
\end{aligned}
$$

Now,

$$
\begin{aligned}
&I_{22}^1(\mathsf{SNR}, A)\\
&\quad= \frac{1}{(1+A)^{r+1}}\int_0^{\infty}\varsigma(\varsigma+\varsigma^*)^{r-1}\exp\Big(-\frac{\varsigma}{1+A}\Big)d\varsigma\\
&\quad= \frac{1}{(1+A)^{r+1}}\exp\Big(\frac{\varsigma^*}{1+A}\Big)\Big[\int_{\varsigma^*}^{\infty}\varsigma^{r}\exp\Big(-\frac{\varsigma}{1+A}\Big)d\varsigma-\varsigma^*\int_{\varsigma^*}^{\infty}\varsigma^{r-1}\exp\Big(-\frac{\varsigma}{1+A}\Big)d\varsigma\Big]\\
&\quad= \frac{1}{(1+A)^{r+1}}\exp\Big(\frac{\varsigma^*}{1+A}\Big)\Big[(1+A)^{r+1}\Gamma(r+1,\frac{\varsigma^*}{1+A})-\varsigma^*(1+A)^r\Gamma(r,\frac{\varsigma^*}{1+A})\Big]\\
&\quad= \frac{1}{(1+A)^{r+1}}\exp\Big(\frac{\varsigma^*}{1+A}\Big)\Big[\Big[r(1+A)^{r+1}-\varsigma^*(1+A)^r\Big]\Gamma(r,\frac{\varsigma^*}{1+A})+(1+A)(\varsigma^*)^r\exp\Big(-\frac{\varsigma^*}{1+A}\Big)\Big]\\
&\quad= \frac{A}{1+A}I_{22}^{1A}(r,\mathsf{SNR}, A),
\end{aligned}
$$

where,

$$
I_{22}^{1A}(r,\mathsf{SNR}, A) = r!\Big[\sum_{j=0}^{r-1}\frac{[\frac{\varsigma^*}{1+A}]^j}{j!}\Big]\Big[1-\frac{\varsigma^*}{r(1+A)}\Big]+\Big[\frac{\varsigma^*}{1+A}\Big]^r.
$$

Since $\frac{\varsigma^*}{1+A}\to 0$ as $A\to\infty$,

$$
\lim_{A\to\infty} I_{22}^{1A}(r,\mathsf{SNR}, A) = r! \qquad (82)
$$

Thus,

$$I_{22}^1(r, \mathsf{SNR}, A) = \frac{Ar!}{1+A}. \tag{83}$$

We now compute $I_{22}^2(\mathsf{SNR}, A)$.

$$\begin{aligned}
& I_{22}^2(\mathsf{SNR}, A) \\
&= \frac{1}{A(1+A)^r} \int_0^\infty (t+t^*)^{r-1} \exp\Big(-\frac{t}{1+A}\Big) \log\Big[1+\exp\Big(-\frac{At}{1+A}\Big)\Big] dt \\
&\le \frac{1}{A(1+A)^r} \int_0^\infty (t+t^*)^{r-1} \exp\Big(-\frac{t}{1+A}\Big) \exp\Big(-\frac{At}{1+A}\Big) dt \\
&= \frac{1}{A(1+A)^r} \exp(t^*) \int_{t^*}^\infty t^{r-1} \exp(-t) dt \\
&= \frac{1}{A(1+A)^r} \exp(t^*) \Gamma(r, t^*) \\
&= \frac{(r-1)!}{A(1+A)^r} \sum_{j=0}^{r-1} \frac{(t^*)^j}{j!}.
\end{aligned}$$

Moreover,

$$\begin{aligned}
& I_{22}^2(\mathsf{SNR}, A) \\
&\ge \frac{1}{A(1+A)^r} \int_0^\infty (t+t^*)^{r-1} \exp\Big(-\frac{t}{1+A}\Big) \exp\Big(-\frac{At}{1+A}\Big) dt \\
&\quad - \frac{1}{2A(1+A)^r} \int_0^\infty (t+t^*)^{r-1} \exp\Big(-\frac{t}{1+A}\Big) \exp\Big(-\frac{2At}{1+A}\Big) dt \\
&= \Big[\frac{(r-1)!}{A(1+A)^r} \sum_{j=0}^{r-1} \frac{(t^*)^j}{j!}\Big] - \frac{1}{2A(1+A)^r} \int_0^\infty (t+t^*)^{r-1} \exp\Big(-\frac{(1+2A)t}{1+A}\Big) dt \\
&= \Big[\frac{(r-1)!}{A(1+A)^r} \sum_{j=0}^{r-1} \frac{(t^*)^j}{j!}\Big] - \frac{1}{2A(1+A)^r} \exp\Big(\frac{(1+2A)t^*}{1+A}\Big) \Big[\frac{1+A}{1+2A}\Big]^r \Gamma(r, \frac{(1+2A)t^*}{1+A}) \\
&= \Big[\frac{(r-1)!}{A(1+A)^r} \sum_{j=0}^{r-1} \frac{(t^*)^j}{j!}\Big] - \frac{(r-1)!}{2A(1+A)^r} \Big[\frac{1+A}{1+2A}\Big]^r \sum_{j=0}^{r-1} \frac{[(\frac{1+2A}{1+A})t^*]^j}{j!}
\end{aligned}$$

Now as $A \to \infty$, $\frac{t^*}{1+A} \to 0$. Therefore,

$$\lim_{A\to\infty} \frac{(r-1)!}{A(1+A)^r} \sum_{j=0}^{r-1} \frac{(t^*)^j}{j!} = 0,$$

$$\lim_{A\to\infty} \Big[\frac{(r-1)!}{A(1+A)^r} \sum_{j=0}^{r-1} \frac{(t^*)^j}{j!}\Big] - \frac{(r-1)!}{2A(1+A)^r} \Big[\frac{1+A}{1+2A}\Big]^r \sum_{j=0}^{r-1} \frac{[(\frac{1+2A}{1+A})t^*]^j}{j!} = 0.$$

As both the upper and lower bounds go to 0, we have

$$\lim_{A\to\infty} I_{22}^2(\mathsf{SNR}, A) = 0. \tag{84}$$

Substituting (83) and (84) in (81), we have

$$\begin{aligned} & I_{22}(\mathsf{SNR}, A) \\ & = \frac{1}{(r-1)!}\left[\frac{1}{A(1+A)^r}\right]^{\frac{1}{A}} \mathsf{SNR}^{1+\frac{1}{A}}\left[\frac{Ar!}{1+A}\right] + o(\mathsf{SNR}^2), \\ & = rA^{-\frac{r+1}{A}}\mathsf{SNR}^{1+\frac{1}{A}} + o(\mathsf{SNR}^2). \end{aligned}$$

Therefore,

$$I_2(\mathsf{SNR}, A) = rA^{-\frac{r+1}{A}}\mathsf{SNR}^{1+\frac{1}{A}} + o(\mathsf{SNR}^2). \tag{85}$$

Substituting (78) and (85) in (71) and (70), we obtain

$$I(\mathbf{x}, \vec{\mathbf{y}}) = r\mathsf{SNR} - r\mathsf{SNR}\frac{\log(1+A)}{A} - rA^{-\frac{r+1}{A}}\mathsf{SNR}^{1+\frac{1}{A}} + o(\mathsf{SNR}^2).$$

Let the capacity of the channel be $C(\mathsf{SNR})$. Since, on-off signaling may not be optimal for the channel, we will denote the highest achievable rate using on-off signaling as $C_{\mathsf{on-off}}(\mathsf{SNR})$. $C_{\mathsf{on-off}}(\mathsf{SNR})$, is given by

$$\begin{aligned} & C_{\mathsf{on-off}}(\mathsf{SNR}) \\ & = \max_A I(\mathbf{x}, \vec{\mathbf{y}}) \\ & = r\mathsf{SNR}[1 - M^*(\mathsf{SNR})] + o(\mathsf{SNR}^2), \end{aligned} \tag{86}$$

where,

$$\begin{aligned} & M^*(\mathsf{SNR}) \\ & = \min_A \left[\frac{\log(1+A)}{A} + A^{-\frac{r+1}{A}}\mathsf{SNR}^{\frac{1}{A}}\right] \\ & = \min_A \left[\frac{\log(A)}{A} + A^{-\frac{r+1}{A}}\mathsf{SNR}^{\frac{1}{A}}\right]. \end{aligned}$$

The last equality holds since $A$ is large. Let us denote

$$M(A, \mathsf{SNR}) = \min_A \left[\frac{\log(A)}{A} + A^{-\frac{r+1}{A}}\mathsf{SNR}^{\frac{1}{A}}\right].$$

We will prove the following theorem to get a lower bound on $M^*(\mathsf{SNR})$.

**Theorem 5**

$$M^*(\mathsf{SNR}) \geq M_L(\mathsf{SNR}) \triangleq \frac{\log\log(\frac{r}{\mathsf{SNR}})}{\log(\frac{r}{\mathsf{SNR}})} \tag{87}$$

*Proof:* We will prove this by contradiction. Let there be an $A_1$ such that the theorem does not hold. Since $\frac{\log(A_1)}{A_1} \geq 0$ and $A_1^{-\frac{r+1}{A_1}} \mathsf{SNR}^{\frac{1}{A_1}} \geq 0$, we have,

$$\frac{\log(A_1)}{A_1} \leq M_L(\mathsf{SNR}), \tag{88}$$

$$A_1^{-\frac{r+1}{A_1}} \mathsf{SNR}^{\frac{1}{A_1}} \leq M_L(\mathsf{SNR}). \tag{89}$$

If (88) holds, we have

$$A_1 \geq \log(\frac{r}{\mathsf{SNR}}).$$

Moreover,

$$\begin{aligned} & A_1^{-\frac{r+1}{A_1}} \mathsf{SNR}^{\frac{1}{A_1}} \\ & \geq A_1^{-\frac{r+1}{A_1}} \left[\frac{\mathsf{SNR}}{r}\right]^{\frac{1}{A_1}} \\ & = \exp(-\frac{(r+1)\log(A_1)}{A_1}) \left[\frac{\mathsf{SNR}}{r}\right]^{-\frac{1}{\log(\frac{\mathsf{SNR}}{r})}} \\ & \geq \exp[-(r+1)M_L(\mathsf{SNR})]e^{-1} \end{aligned}$$

As $\mathsf{SNR} \to 0$, we have

$$\begin{aligned} \exp[-(r+1)M_L(\mathsf{SNR})]e^{-1} & \gg M_L(\mathsf{SNR}), \\ \Rightarrow A_1^{-\frac{r+1}{A_1}} \mathsf{SNR}^{\frac{1}{A_1}} & \gg M_L(\mathsf{SNR}). \end{aligned}$$

This contradicts (89), which completes the proof. □

To get an upper bound for $M^*(\mathsf{SNR})$, we pick a value of $A$. Let

$$A_2 = \frac{\log(\frac{r}{\mathsf{SNR}})}{\log\log(\frac{r}{\mathsf{SNR}})}.$$

Now,

$$M^*(\mathsf{SNR}) \leq \frac{\log(A_2)}{A_2} + A_2^{-\frac{r+1}{A_2}} \mathsf{SNR}^{\frac{1}{A_2}}. \tag{90}$$

We have

$$\begin{aligned} & \frac{\log(A_2)}{A_2} \\ & = \frac{[\log\log(\frac{r}{\mathsf{SNR}}) - \log\log\log(\frac{r}{\mathsf{SNR}})]\log\log(\frac{r}{\mathsf{SNR}})}{\log(\frac{r}{\mathsf{SNR}})} \\ & \leq \frac{[\log\log(\frac{r}{\mathsf{SNR}})]^2}{\log(\frac{r}{\mathsf{SNR}})}, \end{aligned} \tag{91}$$

and,

$$
\begin{aligned}
&A_2^{-\frac{r+1}{A_2}}\mathsf{SNR}^{\frac{1}{A_2}} && (92)\\
&= \left[\frac{r}{A_2^{r+1}}\right]^{\frac{1}{A_2}}\left[\frac{\mathsf{SNR}}{r}\right]^{\frac{1}{A_2}},\\
&\le \left[\max_r \frac{r}{A_2^{r+1}}\right]^{\frac{1}{A_2}}\left[\frac{\mathsf{SNR}}{r}\right]^{\frac{1}{A_2}},\\
&\le \left[\frac{1}{eA_2\log(A_2)}\right]^{\frac{1}{A_2}}\left[\frac{\mathsf{SNR}}{r}\right]^{\frac{1}{A_2}},\\
&\le \left[\frac{\mathsf{SNR}}{r}\right]^{\frac{1}{A_2}} && (93)\\
&= \left[\frac{\mathsf{SNR}}{r}\right]^{-\frac{\log\log(\frac{r}{\mathsf{SNR}})}{\log(\frac{\mathsf{SNR}}{r})}}\\
&= \frac{1}{\log(\frac{r}{\mathsf{SNR}})}. && (94)
\end{aligned}
$$

Equation (93) holds since $A_2 >> 1$ for $\mathsf{SNR} \to 0$, which makes

$$\left[\frac{1}{eA_2\log(A_2)}\right]^{\frac{1}{A_2}} \le [1]^{\frac{1}{A_2}} = 1.$$

Combining (90,91,94), we have

$$M^*(\mathsf{SNR}) \le \frac{[\log\log(\frac{r}{\mathsf{SNR}})]^2+1}{\log(\frac{r}{\mathsf{SNR}})}. \quad (95)$$

Finally, using (86), Theorem 1 and (95), we have

$$
\begin{aligned}
&r\mathsf{SNR} - r\mathsf{SNR}\frac{[\log\log(\frac{r}{\mathsf{SNR}})]^2+1}{\log(\frac{r}{\mathsf{SNR}})} + o(\mathsf{SNR}^2)\\
&\qquad \le C_{\mathsf{on-off}}(\mathsf{SNR}) \le r\mathsf{SNR} - r\mathsf{SNR}\frac{\log\log(\frac{r}{\mathsf{SNR}})}{\log(\frac{r}{\mathsf{SNR}})} + o(\mathsf{SNR}^2). \quad (96)
\end{aligned}
$$

Since on-off signaling may not be optimal

$$C_{\mathsf{on-off}}(\mathsf{SNR}) \le C(\mathsf{SNR}). \quad (97)$$

As conditioning reduces entropy, we can express the input-output mutual information as

$$I(\mathbf{x}, \vec{\mathbf{y}}) \le \sum_{k=1}^{r} I(\mathbf{x}, \mathbf{y}_k). \quad (98)$$

Each term on the right hand side of (98) is maximized by an on-off distribution [13], and we know from [26] that with this distribution, the mutual information $\forall k \in \{1 \dots r\}$ is

$$I(\mathbf{x}, \mathbf{y}_k) \le \mathsf{SNR} - \mathsf{SNR}\frac{\log\log(\frac{1}{\mathsf{SNR}})}{\log(\frac{1}{\mathsf{SNR}})} + o(\mathsf{SNR}^2).$$

Hence, we can upper bound the capacity as

$$C(\mathsf{SNR}) \leq r\mathsf{SNR} - r\mathsf{SNR}\frac{\log\log(\frac{1}{\mathsf{SNR}})}{\log(\frac{1}{\mathsf{SNR}})} + o(\mathsf{SNR}^2).$$

Since,

$$\frac{\log\log(\frac{r}{\mathsf{SNR}})}{\log(\frac{r}{\mathsf{SNR}})} \leq \frac{\log\log(\frac{1}{\mathsf{SNR}})}{\log(\frac{1}{\mathsf{SNR}})},$$

we have

$$C(\mathsf{SNR}) \leq r\mathsf{SNR} - r\mathsf{SNR}\frac{\log\log(\frac{r}{\mathsf{SNR}})}{\log(\frac{r}{\mathsf{SNR}})} + o(\mathsf{SNR}^2). \tag{99}$$

Combining (96, 97, 99), we obtain

$$r\mathsf{SNR} - r\mathsf{SNR}\frac{[\log\log(\frac{r}{\mathsf{SNR}})]^2 + 1}{\log(\frac{r}{\mathsf{SNR}})} + o(\mathsf{SNR}^2) \leq C(\mathsf{SNR}) \leq r\mathsf{SNR} - r\mathsf{SNR}\frac{\log\log(\frac{r}{\mathsf{SNR}})}{\log(\frac{r}{\mathsf{SNR}})} + o(\mathsf{SNR}^2). \tag{100}$$

We now introduce a notation for the approximation that ignores higher order logarithm functions. Let $f(\mathsf{SNR})$ and $g(\mathsf{SNR})$ be functions of $\mathsf{SNR}$. We will denote

$$f(\mathsf{SNR}) \doteq g(\mathsf{SNR}),$$

if

$$\lim_{\mathsf{SNR}\to 0} \frac{\log f(\mathsf{SNR})}{\log g(\mathsf{SNR})} = 1.$$

With this scaling, the inequalities in (100) become equalities and the capacity can be expressed as

$$C(\mathsf{SNR}) = r\mathsf{SNR} - \Delta^{(t,r)}_{\mathsf{i.i.d}}(\mathsf{SNR}).$$

where,

$$\Delta^{(t,r)}_{\mathsf{i.i.d}}(\mathsf{SNR}) \doteq \frac{r\mathsf{SNR}}{\log(\frac{r}{\mathsf{SNR}})}.$$

Moreover, we also see that on-off signaling (69) is capacity achieving for the i.i.d Rayleigh fading MIMO channel in the wideband regime.(Keeping in mind our scaling, which ignores higher order logarithm functions.)